%% file: JG.tex
\documentclass[12pt]{iopart}

\usepackage{bbold}
\usepackage{booktabs}
\usepackage{hyperref}
\hypersetup{colorlinks=true,
		    linkcolor=blue,
		    citecolor=blue,
		    allcolors=blue}

\usepackage{color}
\usepackage{graphics, graphicx}

\usepackage{ifthen, times, comment}
\usepackage{orcidlink}
\usepackage{tikz}
\usetikzlibrary{arrows}
\usetikzlibrary{decorations.pathreplacing,decorations.markings}

\usepackage[percent]{overpic}
\usepackage{wasysym}

\newcommand{\RN}[1]{%
  \textup{\uppercase\expandafter{\romannumeral#1}}%
}
\input tkz.tex
\newcommand{\AGtoDelta}{$0.025$}
\newcommand{\DeltatoAF}{$0.077693$}
\newcommand{\AFtoRS}{$0.380$}
\newcommand{\RStoAFZ}{$0.790$}
\newcommand{\AFZtoFG}{$0.840$}
\newcommand{\FGtoFM}{$1.110$}
\newcommand{\FMtoFMZ}{$1.500$}
\newcommand{\FMZtoYps}{$1.775$}
\newcommand{\YpstoOmega}{$1.820$}
\newcommand{\OmegatoFMXY}{$1.840$}
\newcommand{\FMXYtoAG}{$1.983$}
\newcommand{\AG}{A$\Gamma$}
\newcommand{\FG}{F$\Gamma$}
\newcommand{\RS}{RS}
\newcommand{\FM}{FM}
\newcommand{\FMZ}{FM-Z}
\newcommand{\FMXY}{FM-XY}
\newcommand{\AF}{AF}
\newcommand{\AFZ}{AF-Z}
\newcommand{\Ups}{$\Upsilon$}
\newcommand{\Omg}{$\Omega$}
\newcommand{\dlt}{$\delta$}

\newcommand{\SC}{$SCE$}
\newcommand{\JG}{$J\Gamma$}
\newcommand{\KG}{$K\Gamma$}
\newcommand{\G}{$\Gamma$}

\newcommand{\RSU}{RS$_{U_6}$}

\begin{document}

\title[Eleven Competing Phases of the \JG\ ladder]
{Eleven Competing Phases in the Heisenberg-Gamma (\JG) Ladder}

\author{Sebastien J. Avakian$^1$ and Erik S. S{\o}rensen$^1$\,\orcidlink{0000-0002-5956-1190}}
\address{$^1$Department of Physics \& Astronomy, McMaster University, Hamilton ON L8S 4M1, Canada.}
\ead{avakians@mcmaster.ca,sorensen@mcmaster.ca}


\begin{abstract}
The spin-orbit generated $\Gamma$ interaction is known to induce strong frustration and to be significant in realistic models
of materials. 
To gain an understanding of the possible phases that can arise from this interaction, it is of considerable interest to 
focus on a limited part of parameter space in a quasi one-dimensional model where high precision numerical results can be obtained.
Here we study the Heisenberg-Gamma (\JG) ladder, determining the complete zero temperature phase diagram by analyzing the entanglement spectrum (ES)
and energy susceptibility.
A total of 11 different phases can be identified, among them the well known rung-singlet (RS) phase and 5 other phases, \FM, \FMZ, \FMXY, \AF\ and \AFZ, with conventional long-range magnetic order.
The 3 ferromagnetic phases, \FM, \FMZ\ and \FMXY\ simultaneously have non-zero scalar chirality.
Two other phases, the antiferromagnetic Gamma (\AG) and ferromagnetic Gamma (\FG) phases, have previously been observed in the Kitaev-Gamma ladder, demonstrating that the
\AG-phase is a symmetry protected topological phase (SPT) 
protected by  $TR\times \mathcal{R}_{b}$ symmetry, the product of time-reversal ($TR$) and $\pi$ rotation around the $b$-axis ($\mathcal{R}_{b}$),
while the \FG-phase is related to the RS phase through a local unitary transformation. The 3 remaining phases, \Ups, \Omg\ and \dlt\ show no conventional order, a doubling of the entanglement spectrum
and for the \Ups\ and \Omg-phases a gap is clearly present. The \dlt-phase has a significantly smaller gap and displays incommensurate correlations, with a peak in the static structure factor, $S(k)$ continuously shifting from $k/\pi\mathord{=}2/3$ to $k\mathord{=}\pi$. 
In the \Omg-phase we find pronounced edge-states consistent with a SPT phase protected by the same $TR\times \mathcal{R}_{b}$ symmetry as the \AG-phase. The precise nature of the \Ups\ and \dlt-phases is less clear.
\end{abstract}
\maketitle

\section{Introduction} \label{Intro}
Often when modelling magnetic materials, the interaction terms are assumed not to depend on the direction of the bond, 
in the sense that terms
that can only be distinguished by their spatial orientation are taken to be equivalent.
Significant interest in models where this is not the case and interactions depend on the direction of bonds, have arisen with Kitaev's exact solutions for the ground-state of a simple local Hamiltonian with bond-directional interactions on a honeycomb lattice, the Kitaev Honeycomb model (KHM)~\cite{kitaev2006anyons}. Bond-directional interactions have previously been considered in the wider context of quantum compass (Kugel-Khomskii) models~\cite{Kugel72, Kugel73, Kugel82, Nussinov2015},
however, for the KHM a spin liquid ground-states can rigorously be demonstrated~\cite{kitaev2006anyons}.
The bond-directional Kitaev interaction ($K$) in the KHM is of the Ising type and can be realized in real materials, as demonstrated by
Jackeli et al.~\cite{cjk2010prl}. This has established the class of Kitaev materials~\cite{rau2016review,winter2017review,hermanns2018review,Takagi2019review,Perkinsreview} that are currently being intensely studied.
Among the most promising candidate materials is $\alpha$-RuCl$_3$~\cite{plumb2014prb,banerjee2016proximate,banerjee2018npj}, a material with two-dimensional honeycomb layers. For $\alpha$-RuCl$_3$ there is growing consensus\cite{winter2017review,hermanns2018review,janssen2017model,Maksimov2020} that the Kitaev interaction is ferromagnetic $K\mathord{<}0$, however, other interactions are clearly also present~\cite{rau2014prl,Katukuri2014} with the \G-interaction the largest~\cite{HSKim2016,Ran2017}. On a given a bond with a Kitaev interaction of the form $KS^\gamma S^\gamma$, the \G-interaction takes the form $\Gamma(S^\alpha S^\beta+S^\beta S^\alpha)$ and it is estimated~\cite{winter2017review,hermanns2018review,janssen2017model,Maksimov2020} that $\Gamma\mathord{>}0$ in $\alpha$-RuCl$_3$.
Interaction terms of the usual Heisenberg form with strength $J$ are also believed to be non-negligible, but smaller than the \G-interaction. Several other interaction terms, such as $\Gamma'$, $J_2$ and $J_3$ are sometimes also taken into considerations, but they are believed
to be even smaller in magnitude for $\alpha$-RuCl$_3$, and we do not discuss them here even though they might crucially influence the phase-diagram due to the very high degree of frustration. The phase diagram of $\alpha$-RuCl$_3$ is of significant current interest due
to the experimental observation a magnetically disordered phase under an applied magnetic field~\cite{banerjee2016proximate,banerjee2018npj,baek2017evidence} which has been interpreted as a spin liquid phase~\cite{kasahara2018thermal,Yokoi2021}. The precise nature of this phase is debated~\cite{Lefrancois2022,Bruin2022,Czajka2023,Chern2021} and a full understanding of the complete phase diagram is lacking. Furthermore, it is clear that the complete phase-diagram of the $K$-$J$-$\Gamma$ model of $\alpha$-RuCl$_3$ in a magnetic field
is very complex and extremely challenging to determine precisely~\cite{rau2014prl,Wang2019prl,Yamada2020prb,luo2021prb}. It is therefore very valuable to study the phase diagram of low dimensional versions of this model in a highly restricted part of parameter space where almost exact results can be obtained from state of the art numerical calculations, and here we focus on the Heisenberg Gamma (\JG) in a ladder geometry. 

While the ladder is a highly restrictive geometry, it can still lead to important insights into the possible phases of the full two-dimensional models, and it includes crucial interactions not present in a purely one-dimensional model. We also note that classes of ladder materials exists that have been shown to 
closely model the ladder geometry~\cite{Dagotto1996,Dagotto1999}, so called spin-ladder materials, and one might hope that it will be possible to find
similar materials with bond-directional interactions.
The ladder geometry is also very attractive since almost exact results can be obtained for extremely large systems or directly in the thermodynamic limit, in stark contrast to the two-dimensional lattice where exact diagonalization results
are limited to very small sizes~\cite{rau2014prl,hickey2018visons}. Multi-leg models have been studied~\cite{Gohlke2018prb,Gohlke2020} but systematic studies are challenging.
Kitaev's~\cite{kitaev2006anyons} solutions of the honeycomb model can be extended to include the ladder~\cite{feng2007characterization} but is not applicable when $J\mathord{\neq}0$ or $\Gamma\mathord{\neq}0$. The Kitaev-Heisenberg model in a ladder geometry has been studied
using numerical techniques~\cite{Catuneanu2018ladder,Agrapidis2018,Agrapidis2019} finding 6 distinct phases at zero field as the ratio $J/K$ is varied, in remarkable good agreement with exact diagonalization results for the honeycomb lattice~\cite{cjk2014zigzag}. Likewise, the Kitaev-Gamma ladder has also been investigated~\cite{Gordon2019,sorensen2021prx,AG}, and in this case 8 distinct phases can be identified in zero field versus $\Gamma/K$. The Heisenberg-Gamma (\JG) ladder is relatively less explored, and from the exact diagonalization results
in Ref.~\cite{rau2014prl} the line in the phase diagram of the honeycomb lattice corresponding to the \JG-model appear to only cross a modest number of phases. Here we show that the phase diagram of the \JG-ladder is significant richer, with a total of 11 distinct phases appearing in zero field. In addition to 5 phases, \FM, \FMZ, \FMXY, \AF\ and \AFZ, with conventional long-range magnetic order we observe three previously discussed phases, the \RS, \FG\ and \AG-phases, where the \RS\ and \FG-phases are related
by a local unitary transformation~\cite{ck2015prb}. However, we also find three new potential SPT phases, that we denote \Ups, \Omg\ and \dlt. These 2 phases show no conventional order, a doubling of the entanglement spectrum
and for the \Ups\ and \Omg-phases a relatively clear gap, consistent with SPT behavior.

SPT phases in gapped one-dimensional spin systems can be classified using a projective symmetry 
analysis~\cite{Pollmann2010,Chen2011,Chen2011b}. Usually the site symmetries given by $D_2\mathord{=}\{E,R_x,R_y,R_z\}$ is considered where
$R_x(R_y,R_z)$ is a $\pi$ rotation about the $x(y,z)$ axis. The projective analysis can be extended to ladders~\cite{Liu2012,Chen2015,Ueda2014,Kariyado2015,Ogino2021,Ogino2022} where the additional symmetry $\sigma$, arising from interchanging the legs of the ladder is also included, and the group $D_2\times\sigma$ is considered. It is important to note that a local unitary transformation, $U_6$ exists~\cite{Chaloupka2015hidden}, that maps the Kitaev-Gamma (\KG) ladder to a model with $D_2\times\sigma$ site symmetry. While $\sigma$ is not a good symmetry for the Kitaev-Heisenbergv ($KJ$) ladder, it does have the $D_2$ symmetry. However, neither $D_2$ nor $\sigma$ is a good symmetry for the \JG-ladder and the $U_6$ transformation is not useful. Instead, the $\sigma$ symmetry is replaced by a non-symmorphic symmetry, involving both reflection and translation. In the following, we therefore mainly focus on the time-reversal (TR) symmetry present in the \JG-ladder in zero field. It can therefore be argued that the effect of the \G-interaction is particularly relevant for the \JG-ladder that is
our focus here, due to the significant reduction in the symmetry.

The outline of the paper is as follows. In section \ref{gamma_ladder}, we introduce the \JG-ladder, the geometry and the parametrization. 
The bulk of our results are obtained using DMRG and iDMRG techniques, and in  section~\ref{methods} we discuss the specifics of our numerical methods along with the conventions used. In section~\ref{phasediagram} we present our main results for the phase-diagram of the \JG-ladder, demonstrating the presence of 11 distinct phases. In section~\ref{mag_phases} the magnetically ordered phases are discussed along with the chiral ordering we observe in the ferromagnetic phases. The \RS, \AG\ and \FG-phases are discussed in section~\ref{gamma_phases}. Finally, the three new potential SPT phases, \Ups, \Omg\ and \dlt\ and their classification are discussed in section~\ref{spt_phases} where we also discuss
the projective symmetry analysis of the time-reversal (TR) symmetry.

\section{The \JG-ladder} \label{gamma_ladder}
\begin{figure}
    \centering
    \includegraphics[scale=0.4]{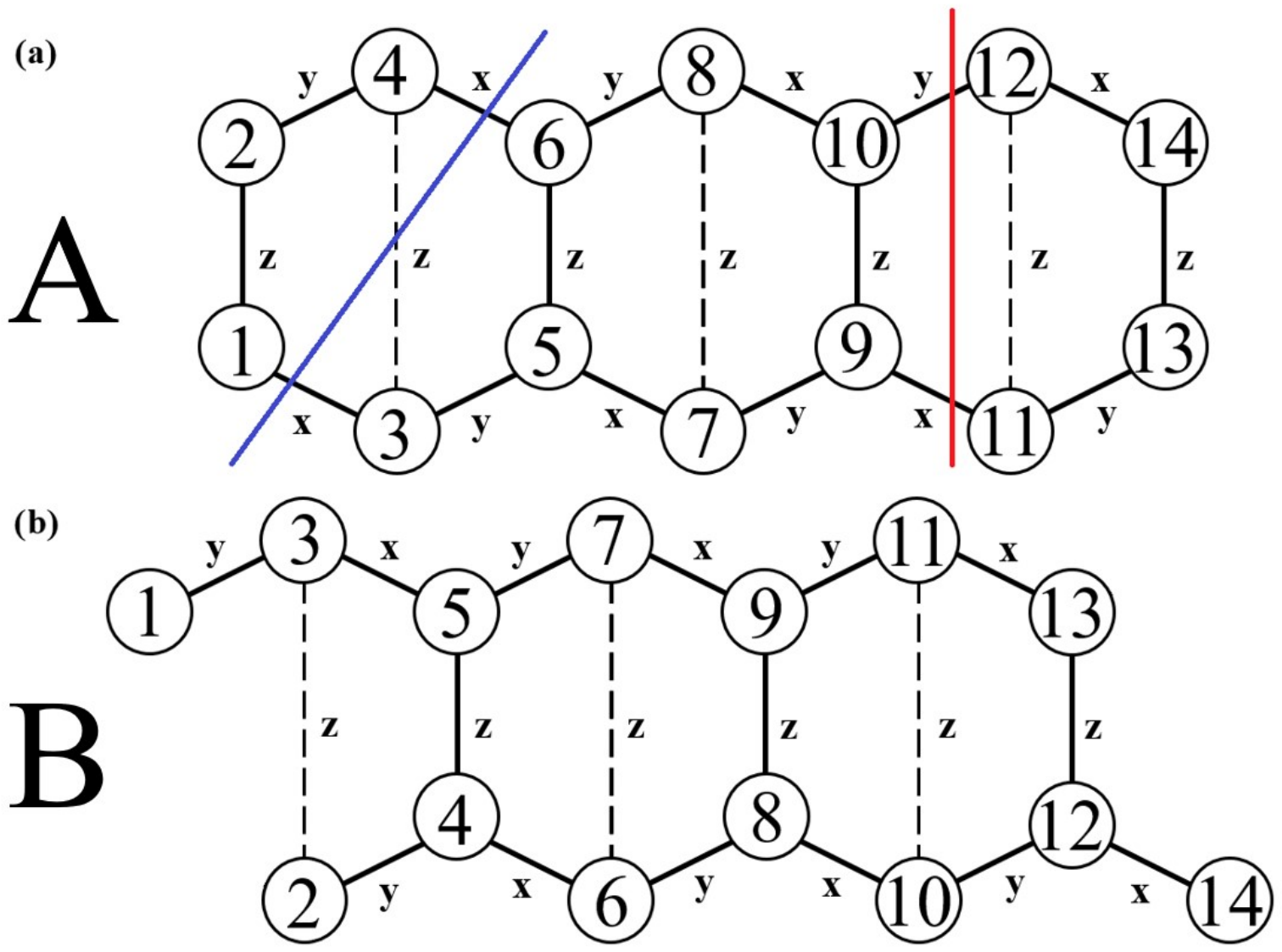}
    \caption{The two clusters of the two leg \JG-ladder with alternating $x$ and $y$ bonds along the legs of the ladder, connected by $z$ bonds along the rungs. Here the $x$, $y$ and $z$ refers to the variation in the \G-interaction. The dashed lines indicate $z$ bonds arising
    from imposing periodic boundary conditions in the perpendicular direction. (a) Cluster A formed from a regular ladder with the red line indicating  a bond cut and the blue line a rung cut, relevant for forming the reduced density matrix. (b) Cluster B formed from cutting the rungs from cluster.} \label{fig:ladder}
\end{figure}
Our focus is on the two leg ladder derived from the honeycomb lattice. 
To get as faithful a representation of the honeycomb lattice as possible,
we consider a small strip of the honeycomb material and create a two leg ladder by ensuring that the bonds that are cut perpendicular to the length of the ladder are paired together, effectively imposing periodic boundary conditions in the perpendicular direction.
This is illustrated in Fig. \ref{fig:ladder} where the dashed bonds arise due to the periodic boundary conditions. We assume these interactions to be of the same strength as the direct coupling between the legs, shown as the solid rungs in Fig. \ref{fig:ladder}.
On each bond of the ladder, we introduce an isotropic Heisenberg interaction of strength $J$.
The second interaction is the \G-interaction, an asymmetric exchange interaction that crucially varies between bonds and is not the same for
every bond. The corresponding Hamiltonian is then
\begin{equation}
    H = \sum_{\langle i,j \rangle} J \mathbf{S}_i \cdot \mathbf{S}_j\  + \sum_{\langle i,j \rangle_\gamma} \Gamma (S^{\alpha}_i S^{\beta}_j + S^{\beta}_i S^{\alpha}_j),\label{eq:HJG}
\end{equation}
where $\langle i,j \rangle_{\gamma}$ denotes the nearest neighbor bond of type $\gamma$. The possible kinds of bonds are $\gamma = x, y, z$ labeling the possible values of $(\alpha,\beta)$ as $(y,z)$, $(x,z)$, and $(x,y)$ respectively. In other words, $\gamma$ labels the missing spin component exchange present in $S_i^{\alpha} S_j^{\beta}$. 

\subsection{Clusters and Boundaries}
The $J$-$\Gamma$ ladder, Eq.~(\ref{eq:HJG}), comprises alternating $x$ and $y$ bonds along the both legs, connected by $z$ bonds along the rungs. The unit cell of the ladder then consists of 4 sites, as shown  in Fig. \ref{fig:ladder}(a) and (b). 
When discussing properties of the model derived from the reduced density matrix of a bipartition of the lattice, it is important to take into account
if the partition cuts a rung or only the legs of the ladder. This is illustrated by the blue and red lines in Fig.~\ref{fig:ladder}(a) showing a rung (blue)
and bond (red) cut respectively. When considering edge-states that may appear in SPT phases, we shall consider finite open segments of the ladder, and it is then crucial to consider how the open boundary conditions are imposed. 
For the ladder, we consider the two possibilities shown in Fig. \ref{fig:ladder}(a) and (b).
We refer to the first (regular) cluster as A and the second (rung cut) cluster as B.  
The degeneracy of the ground-state in a SPT phase strongly depends on whether open or periodic boundary conditions are applied,
however, as we shall see in the following, the degeneracy of the ground-state can also depend on whether cluster A or B is used.

\subsection{Parametrization and connections to known models}
The overall scale for the coupling constants, $J$ and \G\ are not relevant, and it is therefore convenient to
parameterize them in the following way
\begin{equation}
    J = \sin(\phi),\ \Gamma = \cos(\phi).
\end{equation}
The phase space of the model can then be parameterized by the angle $\phi$.
Some points in the phase-diagram correspond to models that have previously been studied in detail.
At $\phi\mathord{=}\pi/2$ the \JG-ladder is simply the antiferromagnetic Heisenberg ladder for which it has been established
that the ground-state is a rung-singlet (RS) state with a sizable gap~\cite{Dagotto1992,Barnes1993,Dagotto1996}. 
Similarly, at $\phi$=$3\pi / 2$ we find the well known ferromagnetic Heisenberg ladder that we expect to show gapless spin wave excitations.

The model with a pure antiferromagnetic \G-interactions occurring at $\phi$=0, has previously been studied in detail~\cite{Gordon2019,sorensen2021prx,AG} and it is known that the model is in a SPT phase with a gap. In addition, a string order parameter has been found~\cite{AG}.
Interestingly, the same antiferromagnetic \G-model on the two-dimensional honeycomb lattice is believed to exhibit a gapless spin liquid phase~\cite{Luo2021npj}
although other scenarios have been discussed~\cite{perkins2017classical,Wang2019prl,Gohlke2020}.

Finally, at $\phi$=$\pi$ we find the ferromagnetic \G-ladder. If we at this point apply the local unitary $U_6$ transformation~\cite{Chaloupka2015hidden}, the \FG-ladder can be mapped to an {\it antiferromagnetic} (AF) spin ladder with nearest neighbor interactions only of the type $S_i^xS_j^x$, $S_i^yS_j^y$ and $S_i^zS_j^z$. Such an AF spin ladder has been shown~\cite{sorensen2021prx} to be in the same phase as the isotropic AF spin ladder with isotropic 
${\bf S}_i\cdot{\bf S_j}$ interactions on each bond which is known to be in the rung-singlet phase, as discussed above. The \FG-phase, of which the $\phi$=$\pi$ is part, can therefore
also be labelled \RSU\ since it is related to the \RS-phase through the local unitary $U_6$ transformation.

\section{Methods} \label{methods}
The main tool used in this analysis is the finite density matrix renormalization group (DMRG)~\cite{White1992a,White1992b,White1993,Schollwock2005,Hallberg2006,Schollwock2011} and its infinite sized version, the infinite density matrix renormalization group (iDMRG)~\cite{McCulloch2008}. The finite size version will be used to obtain the ground state and the next 4 excited states with open boundary conditions (OBC) and periodic boundary conditions (PBC). For the OBC, we mainly use a maximal bond dimension $D\mathord{=} 1000$ and a precision of $\epsilon \mathord{=} 10^{-13}$, while with PBC we typically use $D\mathord{=} 1200$ and $\epsilon \mathord{=} 10^{-11}$. 
To obtain the ground state in the thermodynamic limit, produce the phase diagram, and calculate the bulk correlation functions, we use iDMRG with $D\mathord{=} 1000$ and $\epsilon \mathord{=} 10^{-11}$. 
In order to ensure that we detect all possible phases, the maximum resolution we use is ${\Delta\phi / \pi = 0.001}$.

To detect the quantum critical points (QCP), we use two measures of the ground state wavefunction, the first being the susceptibility of the ground state energy per spin $e_0$ with respect to $\phi$
\begin{equation}
    \chi^e_\phi=-\frac{\partial^2 e_0 }{\partial\phi^2}.
\end{equation}
In finite systems, at a quantum critical point,  $\chi^e$ been shown to scale as~\cite{venuti2007quantum,Schwandt2009,Albuquerque2010}
\begin{equation}
  \chi^e \sim N^{2/\nu-d-z}.
\end{equation}
Here $\nu$ and $z$ are the correlation and dynamical critical exponents and $d$ is the dimension. Hence, only when $2/\nu\mathord{>}d+z$ will $\chi^e$ diverge.
If we assume that $z$=1 and with $d=1$ we find that $\nu<1$ as a condition for a divergence to occur. In addition, the divergence might be very narrow and could be missed if $\Delta\phi$ is not sufficiently small. In principle, when studying systems in the thermodynamic limit with iDMRG, $\chi^e$ should be infinite at the QCP but the finite resolution in $\phi$ will instead show a very sharp spike instead. 
It is therefore very useful to have a complementary way of determining the phase diagram, and
for quasi one-dimensional models this can be obtained from the entanglement spectrum. 
If we cut the ladder across the bond $n$ and form the reduced density matrix $\rho_n$ the entanglement spectrum (ES) can be obtained
from the eigenvalues $\lambda_i$ of $\rho_n$. The eigenvalues change slowly away from a quantum critical point but rapidly near a QCP. 
Sometimes the so-called Schmidt gap~\cite{Chiara2012,Lepori2013,Bayat2014,Gray2018}, the difference between the two largest eigenvalues, is studied, but here we focus
on just the leading eigenvalue $\lambda_1$ which defines the single copy entanglement~\cite{Eisert2005}
\begin{equation}
    SCE = -\ln(\lambda_1).
\end{equation}
When the ground state is in a product state, we must have that $\lambda_1 \mathord{=} 1$ and $\lambda_n \mathord{=} 0$, $\forall n\mathord{>}1$ implying that \SC\ = 0. On the other hand, if our system is not in a product state, $\lambda_1 \mathord{<} 1$, we must have \SC\ $> 0$. 
In the ladder geometry, the only two unique bipartitions are made by either cutting through two leg bonds or through two leg bonds and a rung, as shown in i
Fig.~\ref{fig:ladder}(a). 
As previously outlined, we shall refer to this as a 'bond' cut and a 'rung' cut, respectively.
While either cut can be used for our purposes, we mainly use the rung cut, the blue line in Fig.~\ref{fig:ladder}(a), when studying the \SC.

In order to characterize the magnetic ordering of the phases, 
we study the spin correlation functions $\langle S^{\alpha}_{i}S^{\alpha}_{i+n}\rangle$ as well as the on-site magnetization $\langle S^{\alpha}_{i} \rangle$. In addition, we also study the scalar chirality. 
For any 3 spins $\mathbf{S}\mathord{=}\mathbf{\sigma}/2$ at sites $i$, $j$ and $k$ the scalar chirality is defined as follows,
\begin{equation}
  \kappa = \langle {\bf \sigma}_{i}\cdot \left({\bf \sigma}_{j}\times {\bf \sigma}_{k}\right)\rangle.
\end{equation}
From this definition it seems likely that a non-zero $\kappa$ will be accompanied by more conventional magnetic ordering and in section~\ref{mag_phases}
we discuss the chiral ordering in more detail.

\section{Phase Diagram}\label{phasediagram}
\begin{figure}
    \centering
    \includegraphics[scale=0.3]{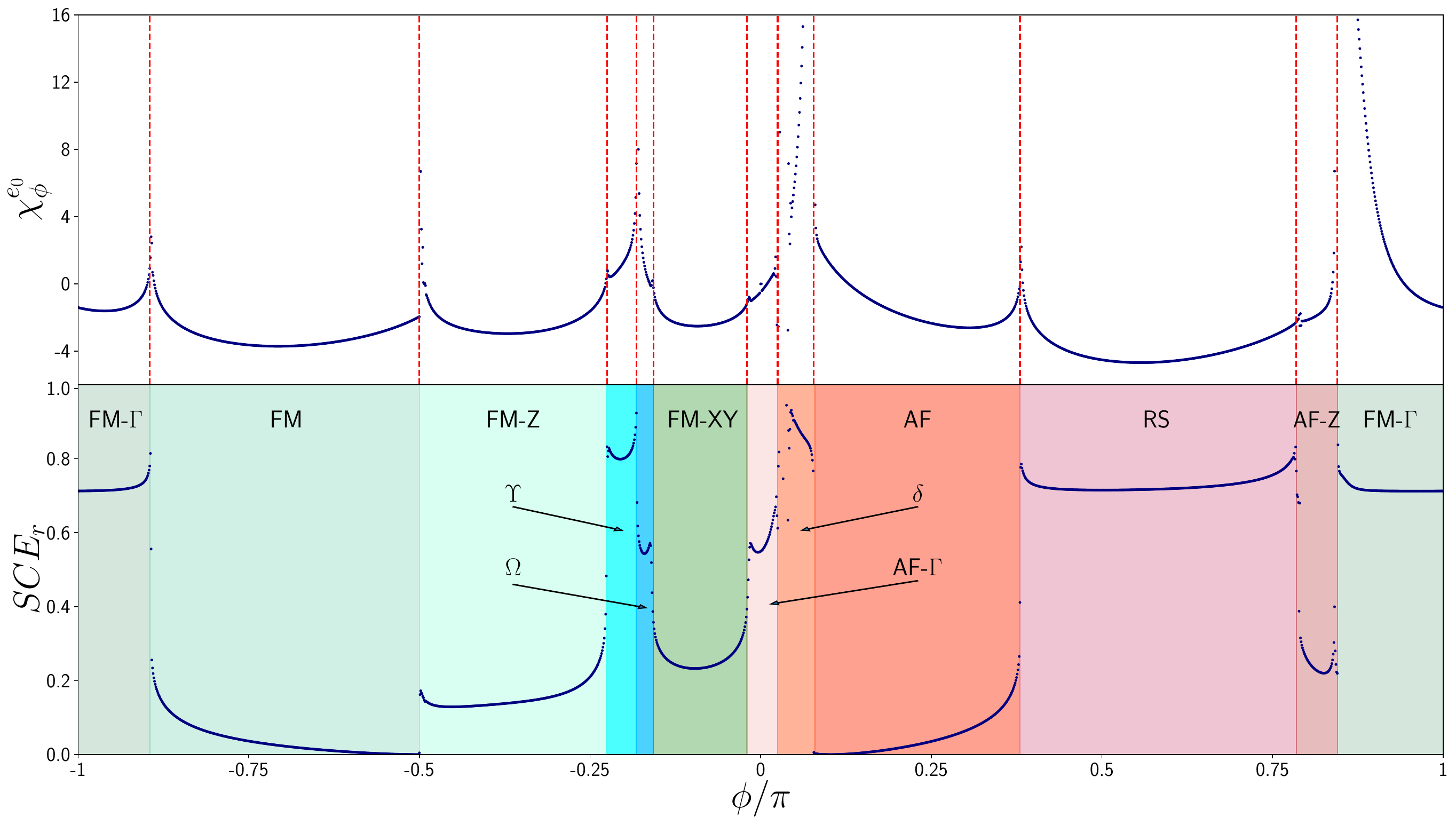}
    \caption{Phase diagram of the \JG-ladder as function of ${\phi}/{\pi}$ from iDMRG with a unit cell of $N$=24 and a resolution ${\Delta\phi} / {\pi} =0.001$. The top panel shows $\chi^e$ while the bottom panel the \SC\ from bond $N/2 -1$ in a rung cut. The dashed lines indicates the quantum critical points.}
    \label{fig:phase_diagram}
\end{figure}
Our main results for the phase diagram are shown in Fig.~\ref{fig:phase_diagram} where we show iDMRG results for $\chi^e$ in the top panel along with results for \SC\ in the bottom panel. The quantum critical points are indicated by the dashed vertical lines. An astonishingly large number of phases is observed, 11 in total. From left to right we label these phases as, \FG, \FM, \FMZ, \Ups, \Omg, \FMXY, \AG, \dlt, \AF, \RS and \AFZ. While most of these phases show some type of magnetic ordering, the \FG, \Ups, \Omg, \AG, \dlt\ and \RS\ phases
do not. We discuss all phases in further detail in the subsequent sections.
At most of the quantum critical points (QCP) we find complete agreement between the divergence in $\chi^e$ and sharp features in the \SC. One exception is the \FMXY\ to \AG\ transition which do not show a clear divergence in $\chi^e$, on the other hand, it is clearly visible in the \SC. This is consistent
with a value of the correlation length exponent $\nu\mathord{>}1$ at this transition. A similar observation can be made about the \RS\ to \AFZ\ transition.
A summary of the results from Fig.~\ref{fig:phase_diagram} can be found  in Table~\ref{summary_of_ph} where the critical values of $\phi$ are listed for all phases along with their characteristics.

\begin{table}
\caption{Summary of the main features of all phases of the \JG-ladder. 
The phase symbol and the critical values of $\phi / \pi$ 
for which the phase exists are listed, as well as the magnetic ordering. The last column indicates the presence of an energy gap in the spectrum in the thermodynamic limit}
\begin{indented}
\lineup
\item[]\begin{tabular}{c | c c c}
\br
 Phase & $\phi_c$ / $\pi $ & Magnetic Ordering & Energy Gap\\ 
 \mr
 \AG\ & \FMXYtoAG\ - \AGtoDelta\ & None & Yes \\ 
 
 $\delta$ & \AGtoDelta\ - \DeltatoAF\ & None & Possibly Gapless \\
 
 AF & \DeltatoAF\ - \AFtoRS\ & AFM & Yes \\
 
 RS & \AFtoRS\ - \RStoAFZ\ & RS & Yes \\
 
 AF-Z & \RStoAFZ\ - \AFZtoFG\ & AFM-Z & Yes \\ 
 
 \FG\ & \AFZtoFG\ - \FGtoFM\ & None & Yes \\
 
 FM & \FGtoFM\ - \FMtoFMZ\ & FM & Yes \\
 
 FM-Z & \FMtoFMZ\ - \FMZtoYps\ & FM-Z & Yes\\
 
 $\Upsilon$ & \FMZtoYps\ - \YpstoOmega\ & None & Yes \\
 
 $\Omega$ & \YpstoOmega\ - \OmegatoFMXY\ & None & Yes  \\
 
 FM-XY & \OmegatoFMXY\ - \FMXYtoAG\ & FM-XY & Yes\\
 \br
\end{tabular}
\end{indented}
\label{summary_of_ph}
\end{table} 

\begin{table}
\caption{Summary of the main features of the potential SPT phases of the \JG-ladder. The $d_{rung}$ and $d_{bond}$ are the degeneracies in the spectrum of the reduced density matrix formed on cluster A or B respectively. 
The $d_{gs}^{OBC_A}$, $d_{gs}^{OBC_B}$, and $d_{gs}^{PBC}$ are the ground state degeneracies in open or periodic boundary conditions with $OBC_A$ and $OBC_B$ referring to cluster A and B respectively.
$\mathcal{O}_\mathrm{TR}^A$ and $\mathcal{O}_\mathrm{TR}^B$ are the projective phase factors under time-reversal (See section~\ref{sec:TR}).
}\label{tab:degeneracy}
\begin{indented}
\lineup
\item[]\begin{tabular}{c | c c c c c | c c}
\br
 Phase & $d_{rung}$ & $d_{leg}$ & $d_{gs}^{OBC_A}$ & $d_{gs}^{OBC_B}$& $d_{gs}^{PBC}$ & $\mathcal{O}_\mathrm{TR}^A$ & $\mathcal{O}_\mathrm{TR}^B$\\ 
\mr
 \AG\  & 1 & 2 & 4 & 1 & 1 & -1 & 1 \\
 
 $\delta$  & 2 & 1 & 1 & 4 & 1 & 1 & -1 \\
 
 
 $\Upsilon$  & 2 & 1 & 1 & 4 & 1 & 1 & -1\\
 
 $\Omega$  & 1 & 2 & 4 & 1 & 1 & -1 & 1\\
 \br
 \end{tabular}
\end{indented}
\end{table}

\subsection{Degeneracies}\label{sec:degs}
A characteristic feature of SPT phases is that edge-states appear under open boundary conditions. A well-known example is the $S$=1 spin chain, where $S$=1/2 edge-states appear~\cite{AKLT87,AKLT88,kennedy1992hidden}
above a four-fold degenerate ground-state with OBC. Another characteristic is a degeneracy of all eigenvalues in the entanglement spectrum~\cite{Pollmann2010,Pollmann2012,Pollmann2012a,Pollmann2012b,Chen2011}.
Such a degeneracy is necessary for non-trivial transformations under the projective symmetry, as we discuss further in section~\ref{spt_phases}. For the determination of the ground-state degeneracy with OBC it is
crucial to distinguish between the two different clusters from Fig.~\ref{fig:ladder}(a) and (b) and we therefore refer to the resulting ground-state degeneracies as $d_{gs}^{OBC_A}$ and $d_{gs}^{OBC_B}$. For the entanglement
spectrum, which we obtain from iDMRG calculations, it is important to distinguish between the rung cut and bond cut discussed above and shown as the blue and red line in Fig.~\ref{fig:ladder}(a). Our results for
these degeneracies for all the potential SPT phases are listed in Table~\ref{tab:degeneracy}. We find in all cases a 4-fold degeneracy of the ground-state using A or B cluster. If the degeneracy is on the A(B) cluster,
the ground-state on the B(A) cluster is non-degenerate. Furthermore, if the 4-fold degeneracy is present on the A(B) cluster, the ES show degeneracy on the bond cut (rung cut) and no degeneracy on the alternate cut.
For completeness, we also list the degeneracy under PBC in Table~\ref{tab:degeneracy}.
These observations are consistent with the presence of SPT phases. Of these 5 phases, the \AG\ and \FG-phases have been studied elsewhere~\cite{Gordon2019,sorensen2021prx,AG} but before analyzing the remaining 3 potential
SPT phases, \Ups, \Omg\ and \dlt, we turn to a discussion of the magnetically ordered phases.

\section{Magnetically Ordered Phases} \label{mag_phases}
\subsection{AF Phases}
\begin{figure}
    \centering
  \includegraphics[scale=0.4]{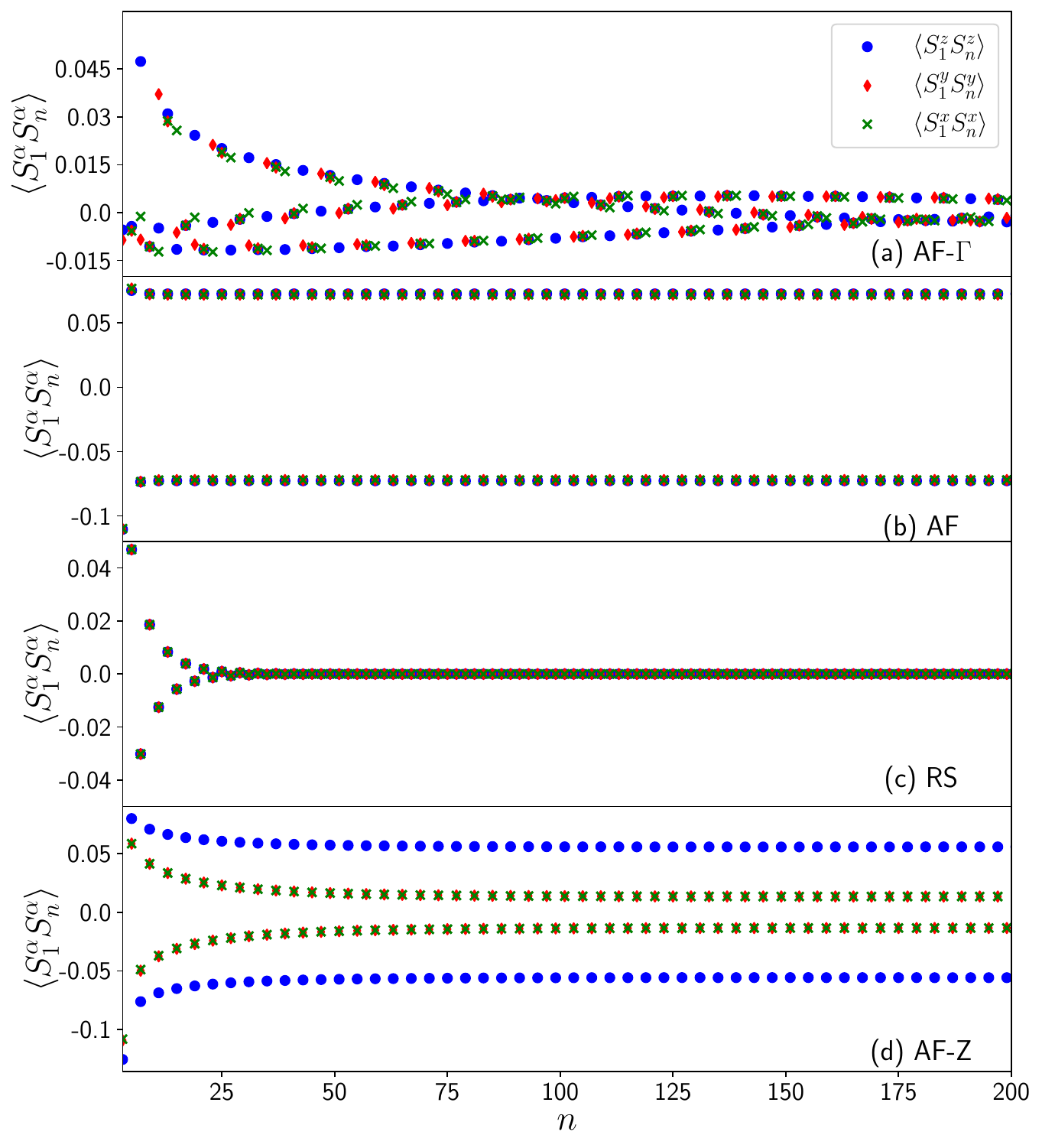}
  \caption{Spin correlation functions $\langle S^\alpha_1S^\alpha_{1+n}\rangle$ along the lower leg of the ladder from $n$=3 to $n$=199, as obtained from iDMRG. Here, $n$ is the site index from Fig.~\ref{fig:ladder}. The \AG-phase at $\phi \mathord{=} 0.064\pi$. (b) The \AF-phase at $\phi\mathord{=}0.249\pi$.
  (c) The \RS-phase at $\phi\mathord{=}0.499\pi$. (d) The \AFZ-phase at $\phi\mathord{=}0.799\pi$.
  }
  \label{fig:corr_af}
\end{figure}  
There are two phases with clear long-range AF magnetic ordering, the \AF\ and \AFZ-phases.
In Fig. \ref{fig:corr_af}(b) and (c) we show results for the spin correlations for each phase.
\begin{itemize}
    \item \AF-phase: For $\phi/\pi \in (\DeltatoAF,\AFtoRS)$, we have the \AF-phase. As can be seen in Fig. \ref{fig:corr_af}(b) the spin correlations are clearly long-range and isotropic
     between the $x$, $y$ and $z$ components. The \G-term is non-zero throughout the \AF-phase, and contrary to what one might expect, the \AF-phase is not gapped. In fact, as we discuss
     in section~\ref{spt_phases}, the correlation length in this phase is rather short, indicating the presence of a well-defined gap.

    \item \AFZ-phase: For $\phi/\pi \in (0.79,0.84)$, the spin correlations look similar to those of the AF phase and are again long-range. However, in this phase the $S^z_1S^z_n$ correlations are larger than the $S^x_1S^x_n$ and $S^y_1S^y_n$ correlations, which are equal,
    as can be seen from our iDMRG results shown in Fig.~\ref{fig:corr_af}(d).
    Hence, we denote this phase the \AFZ-phase. The correlation length is finite, indicating a well-defined gap.

\end{itemize}
\subsection{FM Phases and chiral ordering} 
\begin{figure}
    \centering
  \includegraphics[scale=0.4]{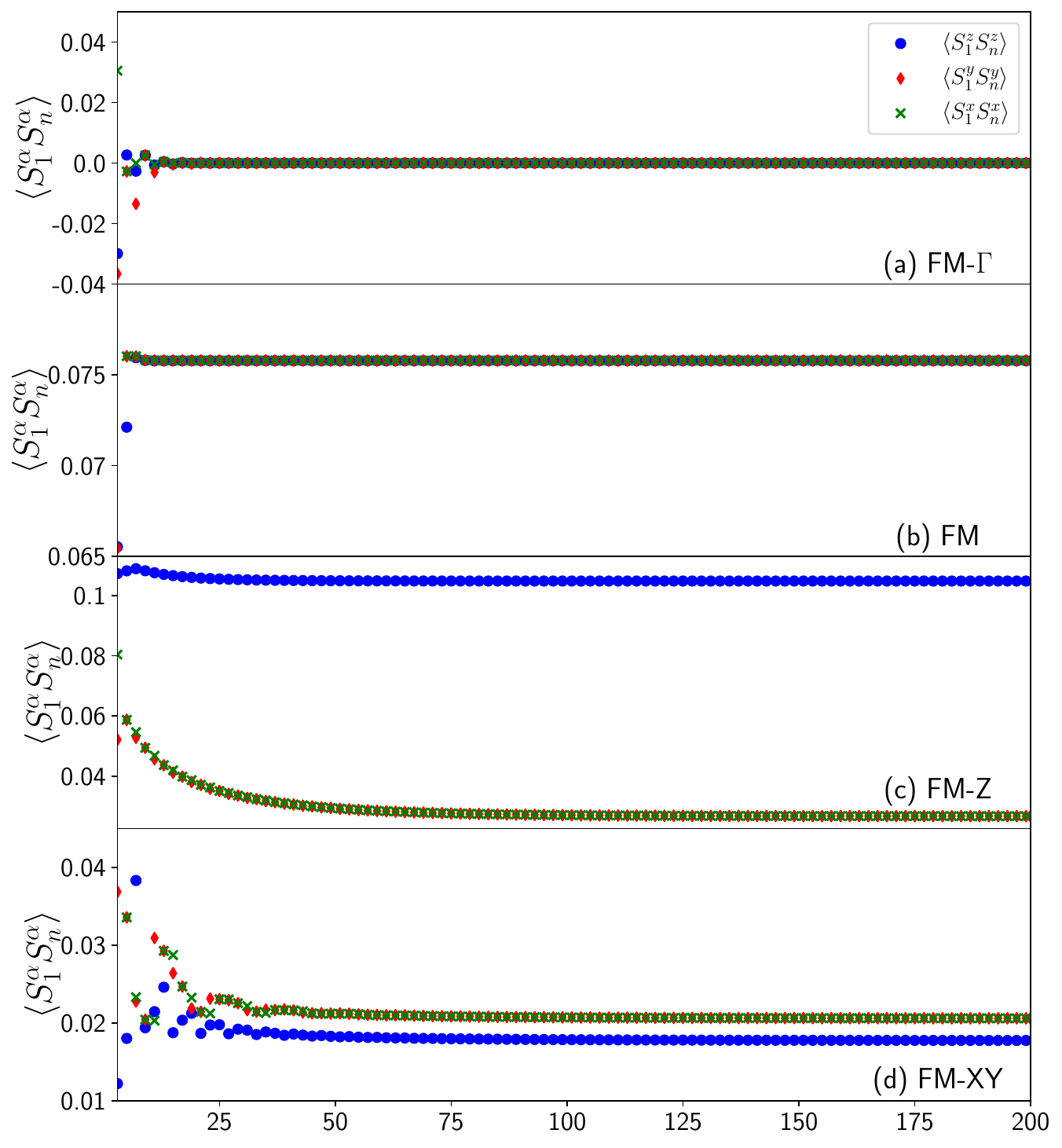}
  \caption{Spin correlation functions $\langle S^\alpha_1S^\alpha_{1+n}\rangle$ as obtained from iDMRG along the lower leg (odd numbered sites) of the ladder starting at $n=3$ and ending at $n=199$. Here, $n$ is the site index from Fig.~\ref{fig:ladder}.
  (a) The \FG-phase at $\phi/\pi$=0.899. (b) The \FM-phase at $\phi/\pi$=1.299. (c) The \FMZ-phase at $\phi/\pi$=1.649. (d) The \FMXY-phase at $\phi/\pi$=1.849.}
  \label{fig:corr_fm}
\end{figure}  
Three of the phases, \FM, \FMZ\ and \FMXY, have long-range ferromagnetic correlations, as can be seen from the results for the spin correlation functions $\langle S^\alpha_1S^\alpha_{1+n}\rangle$ shown in Fig.~\ref{fig:corr_fm}.
\begin{itemize}
    \item \FM-phase: For $\phi/\pi \in (\FGtoFM,\FMtoFMZ)$ the spin correlations shown in Fig.~\ref{fig:corr_fm}(b) show clear long-range ferromagnetic order. Furthermore, all three spin correlation functions appear identical
    and the phase can be identified as an isotropic ferromagnetic phase. As was the case for the AF phases, the \FM-phase is gapped, a fact that we infer from the presence of a relatively short correlation length.

    \item \FMZ-phase: Neighboring the \FM-phase is the \FMZ-phase for $\phi/\pi \in (\FMtoFMZ,\FMZtoYps)$, This phase is similar to the \FM-phase but in the \FMZ-phase the $S^z_1S^z_n$ correlations are larger than the $S^x_1S^x_n$ and $S^y_1S^y_n$, which are equal,
    as illustrated in Fig.~\ref{fig:corr_fm}(c).
    We therefore denote the phase \FMZ.  Similar to the \FM-phase, the \FMZ-phase has a finite correlation length and a gap.

    \item \FMXY-phase: The last ferromagnetic phase is the \FMXY-phase appearing for $\phi/\pi \in (\OmegatoFMXY,\FMXYtoAG)$. Depending on which leg of the ladder is analyzed, the spin correlations have either the $S^x_1S^x_n$ or $S^y_1S^y_n$ correlation marginally larger than the other at small $n$ and then finally equalling each other at larger $n$. Along both legs, the $S^z_1S^z_n$ correlations are smaller than the other two and non-zero. We therefore denote the phase \FMXY. Correlations in the \FMXY-phase are characterized by
    a sizable, but still finite correlation length and therefore a finite gap.
\end{itemize}
It is interesting to note that precisely at $\phi/\pi$=3/2 the \JG-ladder is simply a ferromagnetic Heisenberg ladder, since \G=0. For the FM Heisenberg ladder we would expect gapless spin-wave excitations and an infinite correlation length. This is indeed the case,
since for the \JG-ladder this point corresponds to the transition between the \FM\ and \FMZ\ phases. However, both the \FM\ and \FMZ-phases are gapped, demonstrating the strong effect of the \G-interaction.
\begin{figure}[h!]
  \centering
  \includegraphics[scale=0.4]{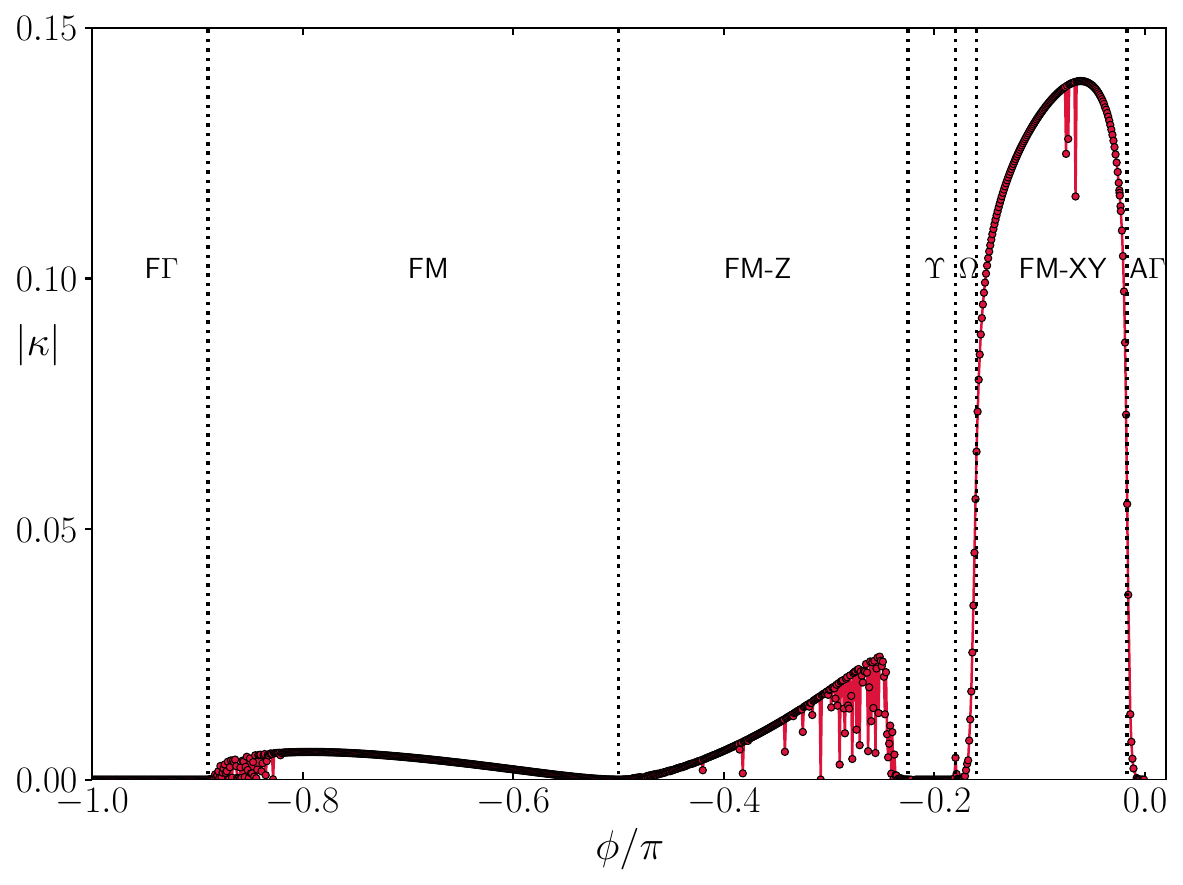}
  \begin{tikzpicture}[scale=1]
    \chrlfmxy{6}
   \end{tikzpicture}
  \caption{Scalar chirality in the \FMXY-phase. Top panel: The scalar chirality $|\kappa|$ versus $\phi/\pi$. Lower panel: An example of the staggered pattern of chirality in the \FMXY-phase at $\phi=1.9\pi$. Each triangle has $|\kappa|\mathord{=}0.133$. 
  The scalar chirality in the \FM\ and \FMZ-phases displays an identical staggering, but $|\kappa|$ is an order of magnitude weaker.}
  \label{fig:chiral}
\end{figure}

From the structure of the \G\ term, it is plausible that it will favor chiral ordering.
Such ordering can accompany conventional magnetic ordering but is sometimes observed even in the
absence of magnetic ordering if three or four-spin interaction terms are present~\cite{Bauer2014,Lauchli2003}.
Since we are considering a ladder, we have to be careful about the handedness of the 3 spins used to measure the chirality
if we want to have a consistent sign convention.
If we label the $i$'th spin on two-legs of the ladder as ${\bf S}_{i,1}$ and ${\bf S}_{i,2}$ where $1$ and $2$ refer
to the bottom-leg and top-leg, respectively, we can then define the scalar chirality by going
around clockwise as follows.
\begin{equation}
  \kappa = \langle {\bf \sigma}_{i,1}\cdot \left({\bf \sigma}_{i,2}\times {\bf \sigma}_{i+1,1}\right)\rangle.
\end{equation}
We get a consistent sign by always going around clockwise. For example, for the upper triangles where we get
$\kappa = \langle {\bf \sigma}_{i,2}\cdot \left({\bf \sigma}_{i+1,2}\times {\bf \sigma}_{i+1,1}\right)\rangle$.
For a pictorial representation,  if the $\kappa$ is positive (negative), we assign blue (red) arrows $i\to j\to k$ for $\kappa=\langle {\bf \sigma}_{i}\cdot \left({\bf \sigma}_{j}\times {\bf \sigma}_{k}\right)\rangle$ 
and all even permutations of $i,j,k$, with clockwise (anti-clockwise) circulation.

The scalar chirality is non-zero in all three ferromagnetic phases, as can be seen from the top panel of Fig.~\ref{fig:chiral}. It is largest in the \FMXY-phase, where a staggered pattern of chirality is observed, alternating in sign between neighboring plaquettes. A sketch of the staggered pattern is shown in the lower panel of Fig.~\ref{fig:chiral}. At $\phi$=$1.9\pi$ in the \FMXY-phase
we find $|\kappa|\mathord{=}0.133$ with $|\kappa|$ going to zero at the quantum critical points of the \FMXY-phase as shown in the top panel of Fig.~\ref{fig:chiral}. 
The same staggered pattern of the chirality is also observed in the \FM\ and \FMZ-phases, but the overall magnitude of $|\kappa|$ is about an order of magnitude
weaker.

\section{\AG, \FG\ and \RS-phases} \label{gamma_phases}
As already discussed briefly in section~\ref{gamma_ladder}, the points $\phi$=0, $\pi/2$ and $\pi$ within the 
\AG, \RS\ and \FG\ phases respectively, (see Fig.~\ref{fig:phase_diagram}), have previously been studied for ladder systems.
Below we list the corresponding phases and their associated properties.

\begin{itemize}
    \item \RS-phase: For $\phi/\pi \in (\AFtoRS,\RStoAFZ)$, we find a rung singlet (RS) phase. This follows from the fact that the phase contains the point $\phi\mathord{=}\pi/2$, where $J$=1 and \G=0, corresponding to the antiferromagnetic Heisenberg ladder. Its ground state is known to be in a disordered  rung singlet phase~\cite{Dagotto1992,Barnes1993}, where the spins on each rung of the ladder are coupled into a spin singlet. 
    This can be confirmed by increasing the Heisenberg coupling of the rungs of the ladder, approaching the product state of rung-singlets.
    The spin correlations for this phase are shown in Fig.~\ref{fig:corr_af}(c) and do not show any long-range magnetic ordering as 
    one would expect. The \RS-phase is gapped, and it has been classified as a trivial SPT phase~\cite{Liu2012}.
    
    \item \FG-phase: This phase extends over the region $\phi/\pi \in (\AFZtoFG,\FGtoFM)$ and includes the point $\phi$=$\pi$ corresponding to the pure ferromagnetic \G\ point with $J$=0, \G=-1. A local unitary transformation, $U_6$, has been found~\cite{Chaloupka2015hidden} that maps the ferromagnetic \G-ladder to a
    model with {\it antiferromagnetic} anisotropic Heisenberg couplings, which is known~\cite{Gordon2019,sorensen2021prx} to be in the same phase as the
    isotropic AF Heisenberg ladder known to be in the \RS-phase. Counterintuitively, the \FG-phase is then simply related to the \RS-phase through the
    $U_6$, a phase that is usually associated with antiferromagnetic interactions, and the \FG-phase is therefore often labelled \RSU.
    As to be expected, spin correlations in the \FG-phase do not show long-range order, as shown in Fig.~\ref{fig:corr_fm}(a).
    The \FG-phase is gapped, with a finite correlation length. Since the \FG-phase is related to the \RS-phase through the local unitary
    $U_6$ transformation, the \FG-phase is also a trivial SPT~\cite{Liu2012}.

    \item \AG-phase: This phase extends over only a small region $\phi/\pi \in (\FMXYtoAG,\AGtoDelta)$ and the pure AF \G\ point at
    $\phi$=0, with $J$=0 and \G=1, has previously been studied in detail~\cite{Gordon2019,sorensen2021prx}. 
    Spin correlations are shown in Fig.~\ref{fig:corr_af}(a) show no long-range magnetic order but a characteristic period 3 variation along the leg
    of the ladder. The phase has a small gap and a sizeable correlation length, with $\xi\sim 41a$ at the pure AF \G\ point.
    The \AG-phase is a SPT phase protected by  $TR\times \mathcal{R}_{b}$ symmetry, the product of time-reversal ($TR$) and $\pi$ rotation around the $b$-axis ($\mathcal{R}_{b}$) and a string-order parameter has been found~\cite{AG}.

\end{itemize}
Above, we have briefly discussed the 5 magnetically ordered phases, \AF, \AFZ, \FM, \FMZ\ and \FMXY\ along with the 3 previously known phases, \RS, \FG\ and
\AG. We now turn to a discussion of the 3 remaining phases, \Ups, \Omg\ and \dlt, all of which show no long-range magnetic order and can be considered as
potential SPT phases.

\section{Potential new SPT phases} \label{spt_phases}
\subsection{Correlation length}
\begin{figure}
    \centering
  \includegraphics[scale=0.5]{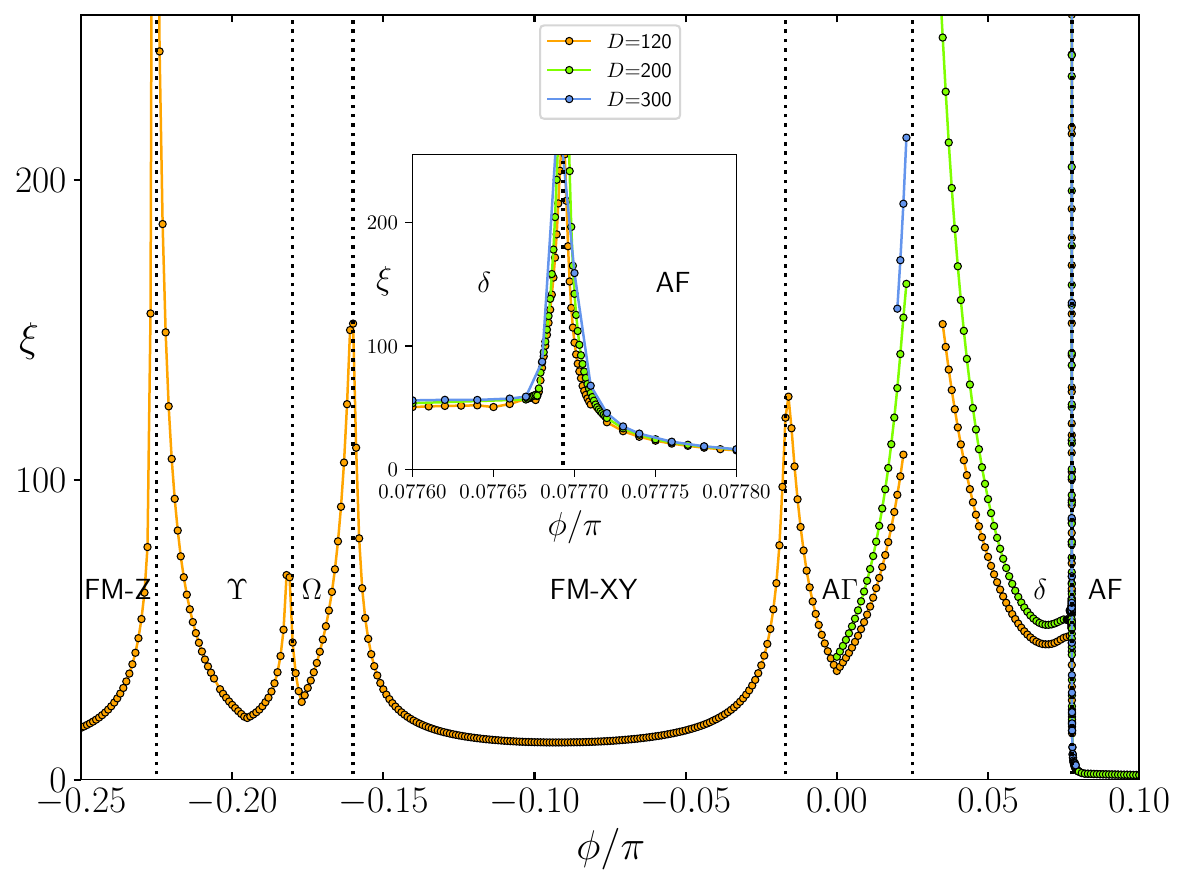}
  \caption{ The correlation length, $\xi$, versus $\phi/\pi$ for $\phi/\pi\in \left[-0.25,0.1\right]$ as obtained from the transfer matrix with bond dimension $D=120,200,300$. The inset shows a close up
  of the $\delta$-AF transition. The dotted lines correspond to the transitions listed in Table~\ref{summary_of_ph}.
    }\label{fig:xi}
\end{figure}  
As is often the case, the most complex part of the phase diagram in Fig.~\ref{fig:phase_diagram} is the proliferation of phases around
the antiferromagnetic \G-point, with $\phi$=$\pi$. We therefore study part of this phase diagram in more detail by explicitly calculating
the correlation length. For translationally invariant matrix product states (MPS), obtained from iDMRG calculations, the transfer matrix can be defined.
For normalized states, the largest eigenvalue of the transfer matrix must be unity, and the second-largest eigenvalue determines the correlation
length through the relation $\xi=-N_c/\ln(|\lambda_2|)$. Here, $N_c$ is the number of sites in the unit cell used to define the transfer matrix.
For quasi one-dimensional systems, the correlation length is related to the gap, $\Delta$, through the relation $\xi=v/\Delta$~\cite{Hastings2006},
with $v$ a characteristic velocity, expected to be ${\cal O}(1)$. If the MPS is obtained with a bond dimension $D$, the transfer matrix is a $D^2\times D^2$ matrix, hindering
calculations of $\xi$ with a very larger bond dimension. However, a significant advantage is that an estimate of the correlation length can be obtained without 
explicit calculations of correlation functions.

In Fig.~\ref{fig:xi} we show results for the correlation length in the region $\phi/\pi\in \left[-0.25,0.1\right]$ from iDMRG calculations with a bond dimension
of $D$=120, 200 and 300. We first note that $\xi$ shows a divergence at all previously noted quantum critical points. Secondly, all phases shown,
\FMZ, \Ups, \Omg, \FMXY, \AG, \dlt\ and \AF, have finite correlation lengths corresponding to a gapped phase. At the mid-point of the potential new SPT phases, we
find approximatively $\xi_\Upsilon\sim 21a$ for the \Ups-phase,
$\xi_\Omega\sim 30a$ for the \Omg-phase and 
$\xi_\delta\sim 57a$ for the \dlt-phase, with $a$ the lattice spacing. Note that, the spin correlation functions shown in Fig.~\ref{fig:corr_af}, \ref{fig:corr_fm} and \ref{fig:corr_spt} are shown along a single leg of the ladder but versus the site index $n$ from Figure~\ref{fig:ladder}. As a function of $n$, they
should therefore decay on a length scale that is given by the correlation length
in Fig.~\ref{fig:xi}, obtained from the transfer matrix. 

\subsection{Spin gap}
\begin{figure}
    \centering
  \includegraphics[scale=0.6]{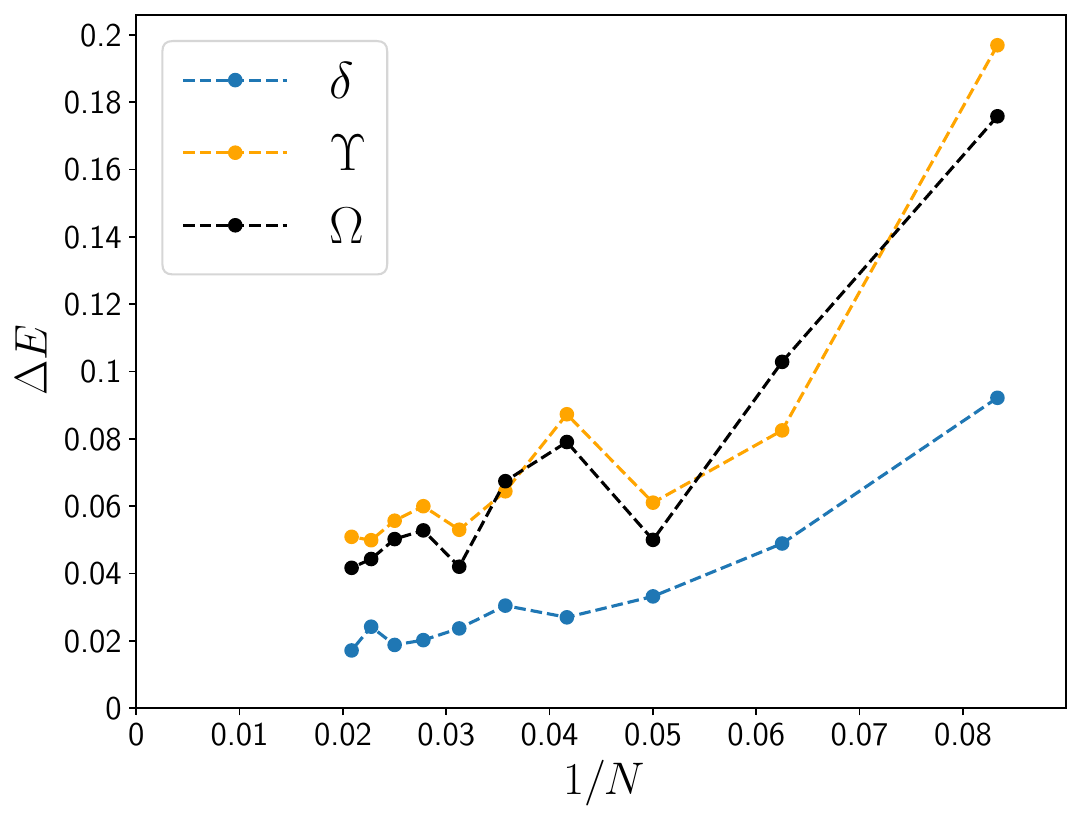}
  \caption{Energy gaps of the \dlt-phase at $\phi$=0.064$\pi$, the \Ups-pase at $\phi$=1.799$\pi$ and the \Omg-phase at $\phi$=1.829$\pi$ 
  between the ground state and the first excited state obtained through finite DMRG with a maximal bond dimension $D$= 1200 and periodic boundary conditions. The system sizes shown correspond to $N$=12 sites to $N$=48 in increments of 4 sites.
    }
  \label{fig:gaps_spt}
\end{figure}
To confirm the presence of a spin gap in the \Ups, \Omg\ and \dlt\
phases, we have explicitly evaluated the gap using exact diagonalization on small systems,
and finite size DMRG calculations with periodic boundary conditions (PBC) on somewhat larger systems. Our results are shown in Fig~\ref{fig:gaps_spt}. 
The dependence on the system size $N$ is not smooth, as one might have expected from the high degree of frustration present in the systems. However, it seems clear that the results for the \Ups and \dlt-phases will converge to a finite small value in the thermodynamic limit, with the gap in the \Ups-phase slightly larger than in the \Omg-phase. This is consistent with our results for $\xi$ that indicate a smaller $\xi$ in the \Ups-phase and therefore likely also a larger gap, when compared to the \Omg-phase if the velocities are assumed the same. The results for the \dlt-phase are more ambiguous, and it seems possible that the gap will tend to zero as $N\to\infty$. However, our results for the $\xi$ in the \dlt-phase close to the \dlt-\AF\ transition are very stable and only show
a small dependence on the bond dimension $D$. At $\phi$=0.077$\pi$ we have the previously quoted value of $\xi\sim 57a$ obtained with $D$=300, $\xi\sim 53a$ ($D$=200) and $\xi\sim 48a$ ($D$=120). 
Although these result indicate a sizable correlation length of $\xi\sim 62a$ as $D\to\infty$, the calculations are very stable, excluding the possibility of a correlation length diverging with $D$ and lending strong support to the
presence of a small but finite gap in the \dlt-phase. It would be interesting to explore the alternative scenario of a gapless \dlt\ phase
by studying the spin stiffness~\cite{Laflorencie2001} in this phase. 

\subsection{Spin correlation functions}
\begin{figure}
    \centering
  \includegraphics[scale=0.4]{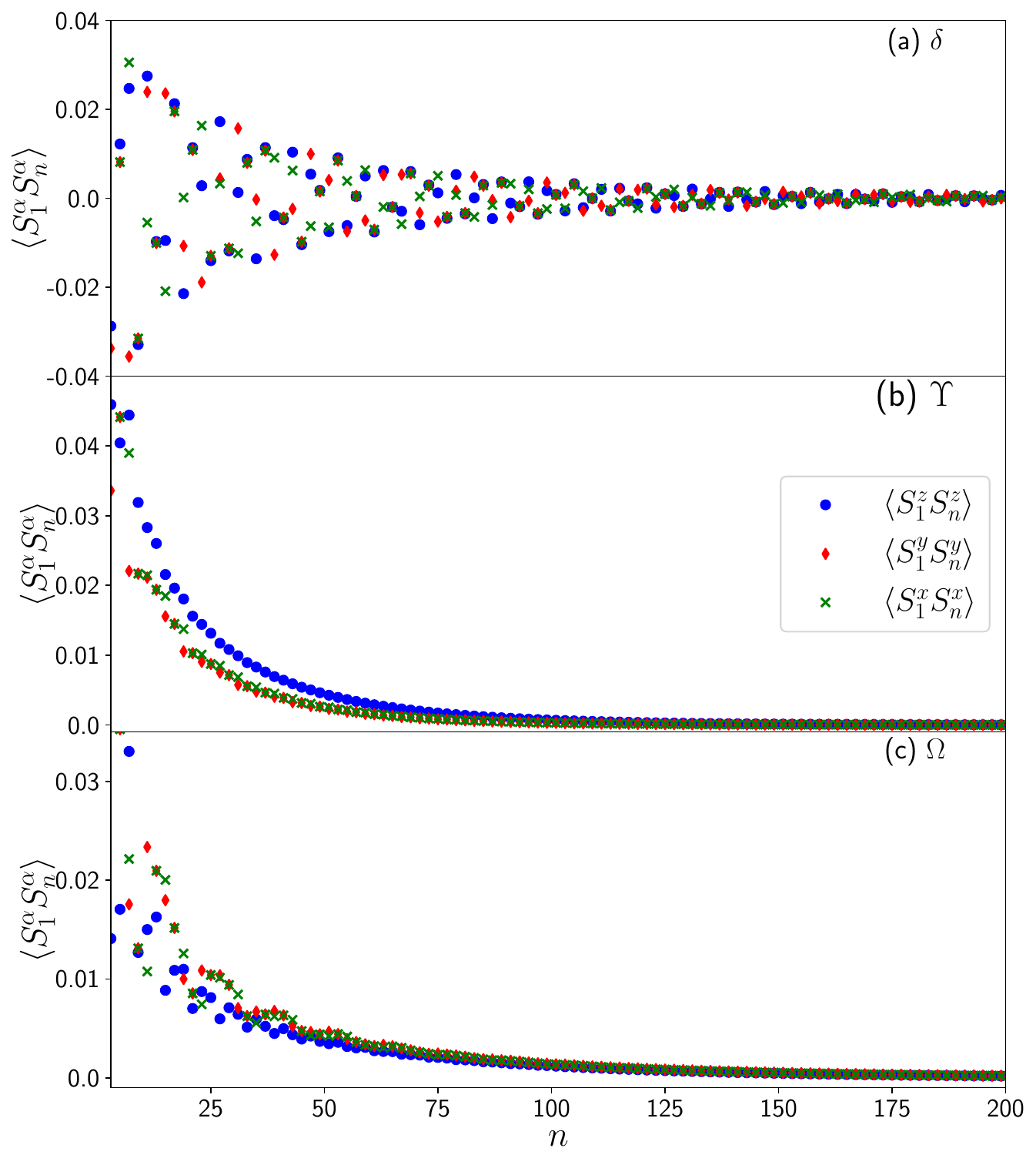}
  \caption{Spin correlation functions, $\langle S^\alpha_1S^\alpha_{1+n}\rangle$ of the potential SPT phases versus $n$, the site index from Fig.~\ref{fig:ladder}.
  Results are for the (a) \dlt-phase at $\phi$=0.064$\pi$, (b) the \Ups-phase at $\phi$=1.799$\pi$ and (c) the \Omg-phase at $\phi$=1.829$\pi$ as obtained from iDMRG.
  Results are shown for correlations along the first leg of the ladder, starting at $n=3$ and ending at $n=199$.
    }
  \label{fig:corr_spt}
\end{figure}  
The spin correlation functions for the three phases, \Ups, \Omg\ and \dlt\ are shown in Fig.~\ref{fig:corr_spt} as obtained from iDMRG calculations.
While the \Ups\ and \Omg-phases show largely ferromagnetic correlations, the \dlt-phase correlations are intermittently negative, showing a more antiferromagnetic nature. However, in all 3 phases, the spin correlation functions quickly approach zero. No long range magnetic order is observed.
As can be seen in Fig.~\ref{fig:chiral} there are no chiral correlations in the \Ups\ and \Omg\ phases, and we have verified that the same is the case
for the \dlt\ phase. The three phases therefore appear to have no discernible order, consistent with the phases being gapped SPT phases.

\subsection{Projective symmetry analysis of time reversal}\label{sec:TR}
\begin{figure}
    \centering
    \includegraphics[scale=0.6]{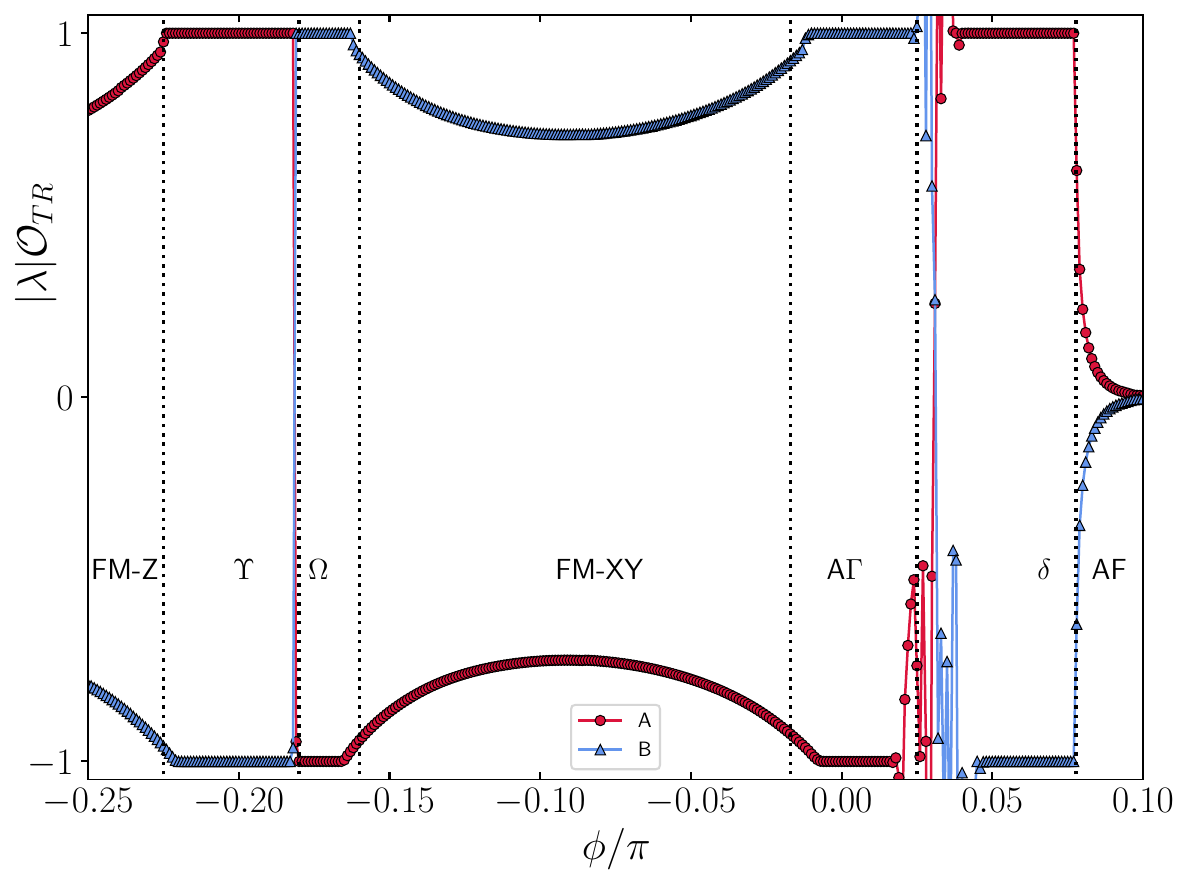}
    \caption{The time-reversal phase factor $|\lambda|\mathcal{O}_\mathrm{TR}$ versus $\phi/\pi$ as obtained from iDMRG, with $|\lambda|$ the leading eigenvalue of the generalized transfer matrix (\ref{eq:transfer}). The red circles correspond to cluster A from Fig.~\ref{fig:ladder}(a) and the blue triangles to cluster B from Fig.~\ref{fig:ladder}(b).
    }
    \label{fig:TR}
\end{figure}
While the usual Landau classification of phases do not distinguish between SPT phases, 
it is possible to develop a classification of such phases based on a projective symmetry analysis~\cite{Gu2009,Chen2011,Liu2011,Chen2011b,Liu2011b,Chen2013}.
This classification takes it starting point in the MPS form of the wave functions for gapped SPT phases where a crucial degeneracy of the entire entanglement spectrum was noted~\cite{Pollmann2010,Pollmann2012a}.

To proceed, one writes the MPS wavefunction in its canonical form~\cite{Vidal2003b,Vidal2007,PerezGarcia2007,Orus2008}:
\begin{equation}
    |\Psi\rangle = \sum_{j_1,\ldots,j_N}\Gamma^{[1]}_{j_1}\Lambda^{[2]}\Gamma^{[2]}_{j_2}\ldots \Lambda^{[N]}\Gamma^{[N]}_{j_N} |j_1,\ldots,j_N\rangle,
\end{equation}
where the $\Gamma^{[n]}_{j_n}$ are complex matrices and the $\Gamma^{[n]}$, real, positive, square diagonal matrices.
If we consider infinite systems with translational symmetry from the perspective of iDMRG, the set of matrices on any unit cell becomes the same 
$\Gamma^{[n]}_{j}$=$\Gamma_{j}$, $\Gamma^{[n]}$ = $\Gamma$ for all $n$, although they may vary within the unit cell.
We now consider a site symmetry operation $g$. In the spin basis this symmetry operation will be  represented by a unitary matrix, $\Sigma_{jj'}(g)$.
One can then establish~\cite{PerezGarcia2008,Pollmann2010} that the $\Gamma_j$ matrices of bond dimension $D$, must transform as~\cite{Pollmann2010,Pollmann2012b}:
\begin{equation}
   \sum_{j'} \Sigma_{jj'}(g)\Gamma_{j'} = e^{i\theta}U^\dagger(g) \Gamma_j U(g), 
\end{equation}
Here, $e^{i\theta}$ is a phase factor, and  the unitary matrices $U(g)$ commute with the $\Gamma$ matrices, and form a $D$-dimensional projective representation of the symmetry group of the wave-function.
Exploiting the full machinery of the MPS formulation, one can show that the $U(g)$ matrices be determined from the unique eigenvector of the generalized transfer matrix with eigenvalue $|\lambda|$=1~\cite{Pollmann2010,Pollmann2012b}.
The generalized transfer matrix is here defined as:
\begin{equation}
T^\Sigma_{\alpha\alpha';\beta\beta'}=\sum_j\left(\sum_{j'}\Sigma_{jj'}\Gamma_{j',\alpha\beta}\right)(\Gamma_{j,\alpha'\beta'})^*\Lambda_{\beta}\Lambda_{\beta'},\label{eq:transfer}
\end{equation}
and it is therefore possible to determine the $U(g)$ matrices numerically once the ground-state wavefunction has been determined in a translationally invariant MPS form.
The projective representation is reflected in the fact that if
$\Sigma(g)\Sigma(h)$=$\Sigma(gh)$, then 
\begin{equation}
   U(g)U(h)=e^{i\phi(g,h)}U(gh),
\end{equation}
where the phases $\phi(g,h)$ are characteristic of the topological phase.

For the \JG-ladder there are few site symmetries and the ladder does not satisfy $D_2$ symmetry, nor is it symmetric with respect to interchanging the legs, $\sigma$. We therefore focus
only on time-reversal $(TR)$, defined by 
$\Gamma_j\to \sum_{j'}\left[e^{i\pi S^y}\right]_{jj'}\Gamma^{\star}_{j'}$, 
with $\star$ denoting complex conjugation. In this case, it can be established that~\cite{Pollmann2010} $U_\mathrm{TR}U_\mathrm{TR}^\star$=$e^{i\phi(TR,TR)}\mathbb{1}$ from which it follows that
$\phi(TR,TR)$=0 or $\pi$. If $\phi(TR,TR)$=$\pi$ we see that $U_\mathrm{TR}$ is an antisymmetric matrix and one can then show~\cite{Pollmann2010} that this is only possible if all eigenvalues of the entanglement
spectrum have even multiplicity, thereby linking the non-trivial value of $e^{i\phi(TR,TR)}$=-1 to the degeneracy of the entanglement spectrum. Furthermore, if the largest eigenvalue of the generalized transfer matrix is smaller than one, $|\lambda|\mathord{<}1$,
then time-reversal is not a good symmetry of the MPS representing the phase.

For time reversal, the phase factor $e^{i\phi(TR,TR)}$ can be extracted by defining
\cite{Pollmann2012b}:
\begin{equation}
\mathcal{O}_\mathrm{TR}\equiv\frac{1}{D}\Tr\left( U_\mathrm{TR}U_\mathrm{TR}^\star \right),\label{eq:OTR}
\end{equation}
with the $D\times D$ matrices $U_\mathrm{TR}$ extracted numerically from the generalized transfer matrix (\ref{eq:transfer}).
For instance, for the $S$=1 spin chain in the Haldane phase,
one finds $\mathcal{O}_\mathrm{TR}\mathord{=}-1$~\cite{Pollmann2010,Pollmann2012a}. In Fig.~\ref{fig:TR} we show results for $|\lambda|\mathcal{O}_\mathrm{TR}$ versus $\phi/\pi$ for the two different clusters
A and B from Fig.~\ref{fig:ladder}. We multiply with $|\lambda|$, the leading eigenvalue of the generalized transfer matrix so that one may immediately see when a phase does not respect the TR symmetry which 
should be the case for the magnetically ordered phases. This is clearly the case for the \FMZ, \FMXY\ and \AF\ phases in Fig.~\ref{fig:TR} where $|\lambda|\mathcal{O}_\mathrm{TR}$ quickly deviates from $\pm$1. In 
section~\ref{sec:degs} we discussed the degeneracy of the ES in the different phases for both cluster A and B. As is clear from Fig.~\ref{fig:TR}, the non-trivial phase factor $\mathcal{O}_\mathrm{TR}\mathord{=}-1$
follows the ES degeneracy and jumps between cluster A and B precisely at the quantum critical points. 
A summary of the results for $\mathcal{O}^{A,B}_\mathrm{TR}$ for the different phases are included in table~\ref{tab:degeneracy}.
Fig.~\ref{fig:TR} shows that if the corresponding cluster is selected, the \Ups, \Omg, and \dlt\ phases all transform non-trivially under TR,
as a non-trivial SPT phase. However, we also note that the \RS-phase has $\mathcal{O}^{A}_\mathrm{TR}\mathord{=}1$ and $\mathcal{O}^{B}_\mathrm{TR}\mathord{=}-1$, and this phase is known to be a trivial SPT phase~\cite{Liu2012}. 
For a definite classification of the
\Ups, \Omg, and \dlt\ phases as non-trivial SPT phases, a further analysis is therefore needed. 

For a more complete picture, it is therefore interesting to study the non-symmorphic symmetry, $\sigma\times\mathrm{tr}(1)$ where $\sigma$ is
the aforementioned operator that interchanges the legs of the ladder while $\mathrm{tr}(1)$ is a translation by 1 lattice spacing in the direction {\it along} the
leg of the ladder. This is a symmetry of the Hamiltonian, Eq.~(\ref{eq:HJG}), but, as far as we can tell, not a symmetry of any of
the \Ups, \Omg, and \dlt\ phases. Still, these phases could be protected by other symmetries that we have not been able to analyze.

\subsection{Uniform field as an active operator}
\begin{figure}
    \centering
    \includegraphics[scale=0.6]{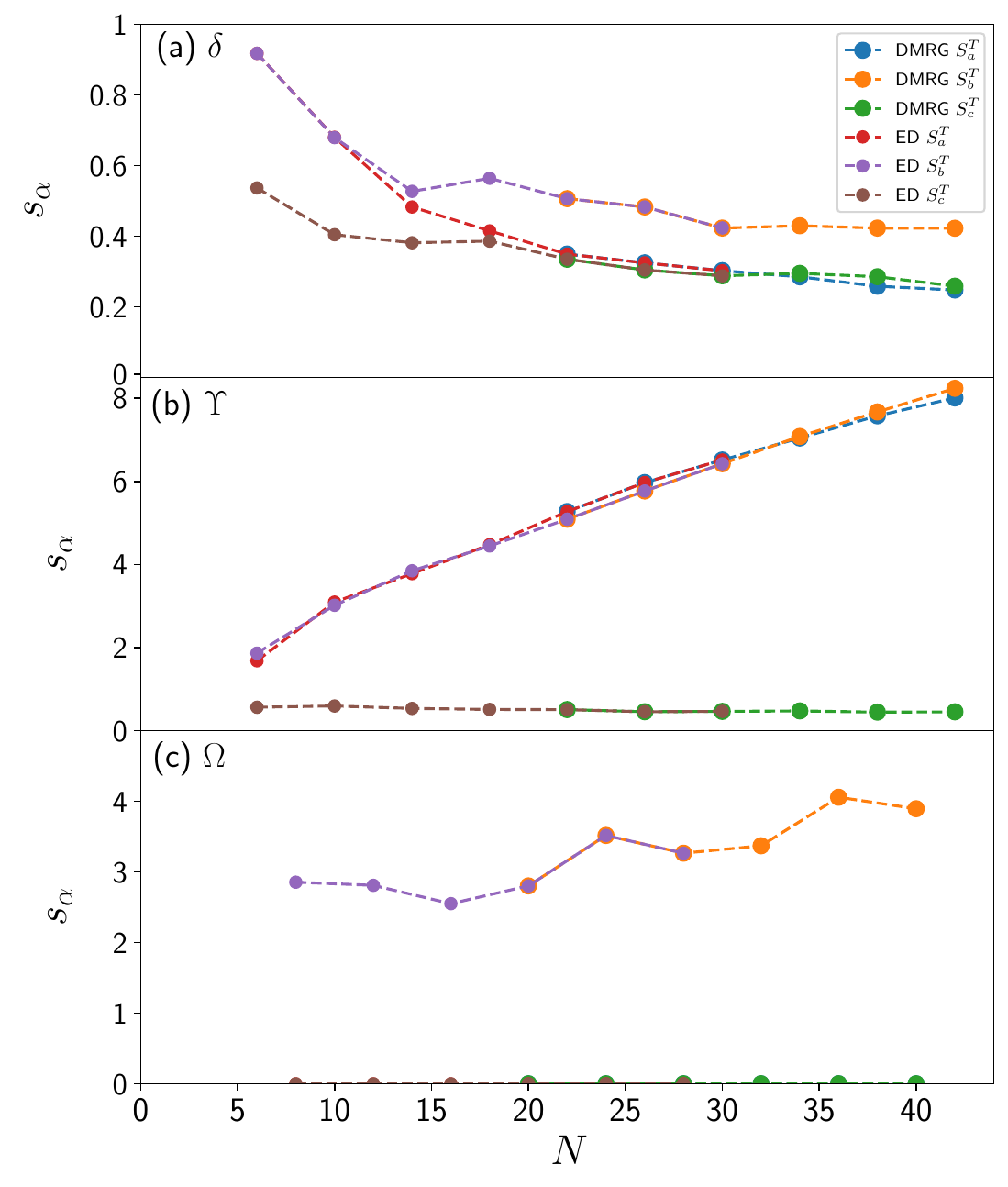}
    \caption{Leading eigenvalue $s_\alpha$ of the total magnetization $S^{T}_{\alpha}$ along axis $\alpha$, with $\alpha=a, b, c$, versus cluster size $N$. (a) $\delta$-phase at $\phi \mathord{=} 0.064\pi$ with cluster B. (b) $\Upsilon$-phase at $\phi \mathord{=} 1.799\pi$ with cluster B.
    (c) $\Omega$-phase at $\phi \mathord{=} 1.829\pi$ with cluster A. For all size until $N=28$, $s$ is obtained with exact diagonalization (in blue, orange, and green) while the remaining larger sizes are obtained with finite DMRG (red, purple, and brown).}
    \label{fig:salpha}
\end{figure}
Each of the three phases, \Ups, \Omg, and \dlt\ show a 4-fold degenerate ground-state. It is of considerable interest to determine what perturbations will split the degeneracy between these four states, the so-called active operators~\cite{Chen2011b,Liu2012,Liu2011,Liu2011b}.
The \JG-ladder does not have the usual site symmetries associated with 180$^\circ$ rotation about the $x(y,z)$ axis. However, it does
possess the previously mentioned $\sigma\times\mathrm{tr}(1)$ symmetry. To study the active operators, we therefore first consider
the behavior of the operator $S^T_\alpha\mathrm{=}\sum_iS_i^\alpha$ within the manifold of the 4 ground-states that we label $|1\rangle, |2\rangle,|3\rangle$ and $|4\rangle$. We note that these operators do not break the $\sigma\times\mathrm{tr}(1)$ symmetry. We must have $\langle i|S^T_\alpha|i\rangle$=0 $\forall i$, since the \Ups, \Omg, and \dlt\ phases
are not magnetically ordered. However, $[S^T_\alpha,H]\neq 0$ so we can diagonalize the matrix $\langle i|S^T_\alpha|j\rangle$ and study
the eigenvalues. A non-zero $s_\alpha$ indicates that the $S^T_\alpha$ operator is active, splitting the states.
In the present case, 
with $\alpha=x,y,z$, all 4 eigenvalues are sometimes non-zero, which is difficult to interpret. However, given the underlying honeycomb lattice, it is natural to instead study
the eigenvalues of $S^T_\alpha$ with $\alpha=a,b,c$, the axis of the honeycomb lattice. Here, $a$ is a unit vector in the $[11\bar 2]$ direction,
$b$ in the $[1\bar{1}0]$ direction and $c$ in the $[111]$ direction. In this case, the results are of the much simpler form $(s_\alpha,-s_\alpha,0,0)$, closely resembling what one finds for the $S$=1 spin chain in the Haldane phase where the 4 ground-states correspond to two free $S$=1/2 excitations at each end, yielding $(1,-1,0,0)$ for $S^T_{x,y,z}$. Results for $s_\alpha$ with $\alpha=a,b,c$ versus system size, $N$, for the \Ups, \Omg, and \dlt\ phases
are shown in Fig.~\ref{fig:salpha}, in each case with cluster A or B from Fig.~\ref{fig:ladder} yielding the 4 degenerate ground-states. For finite systems, the 4 states are not completely degenerate but split by a small amount, decreasing with $N$. Some variation with $N$ is therefore to be expected. In addition, given the results from Fig.~\ref{fig:xi}, showing a large correlation length in all three phases, it is natural to expect that rather large system sizes are needed to see a clear separation of any states localized at the end of the open segments. From studies of the edge excitations in the $S$=1 chains it is known that these excitations fall off as $\exp(-x/\xi)$
from the end of the chain, with $\xi$ the bulk correlation length~\cite{Sorensen1994}, extending far into the chain as the borders of the Haldane phase are approached~\cite{PMS1998}. The clearest results are obtained for the \Omg\ phase, where the results in Fig.~\ref{fig:salpha}(c) show that there is no response to a field in the $a$ and $c$ directions. Furthermore, $s_b$ seems to stabilize around a value $s_b\sim 3-4$, consistent with well-defined edge states. This is the same behavior observed in the \AG\ phase
which was interpreted as a SPT phase protected by $TR\times \mathcal{R}_{b}$ symmetry, the product of time-reversal ($TR$) and $\pi$ rotation around the $b$-axis ($\mathcal{R}_{b}$)~\cite{AG}.

The results for the \dlt\ phase, shown in Fig.~\ref{fig:salpha}(a) are more difficult to interpret. With the limited size available, it seems possible that all $s_\alpha$ could attain a finite small value as $N\to\infty$, or $s_a$ and $s_c$ could reach zero with $s_b$ finite, or all could go to zero. The \dlt\ phase has the largest correlation length of the three phases, and values of $N$ beyond what we have been able to reach are needed to resolve this. 

For the \Ups\ phase, shown in Fig.~\ref{fig:salpha}(b), $s_c$ quickly reach a small finite value $s_c\sim 0.5$, consistent with $S^T_c$ being an active operator. However, surprisingly, $s_a$ and $s_b$ increase with $N$ out to the largest value of $N$. This is not consistent
with $S^T_a$ and $S^T_b$ being active operators. 
We now turn to a brief description of some specific results for the three potential SPT phases.

\subsection{\Omg\ phase}
\begin{figure}
    \centering
    \includegraphics[scale=0.6]{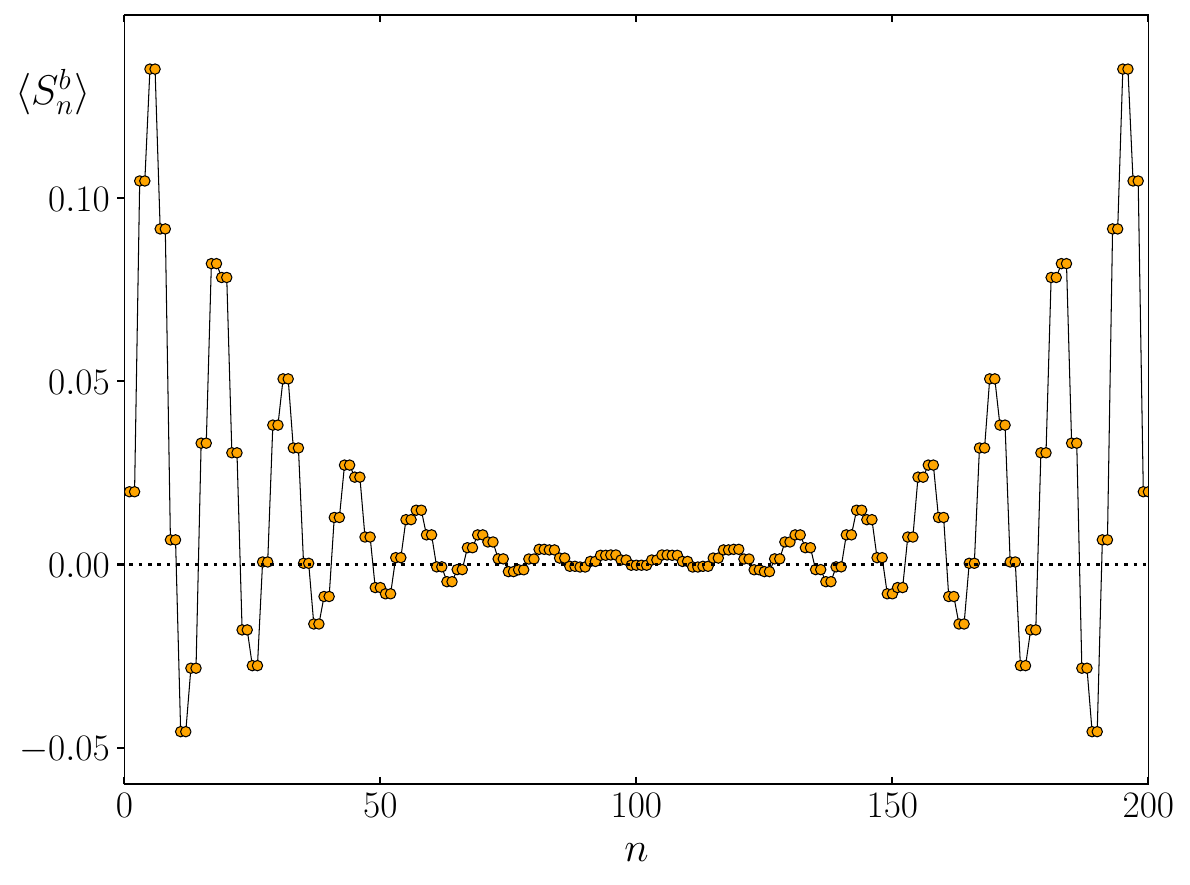}
    \caption{Magnetization $\langle S^b_{n}\rangle$ from finite DMRG calculations along the $b=[1\bar{1}0]$ direction with a uniform field of $h_b=10^{-5}$ applied in the $b$ direction at every site.
    Results are shown for a $N=200$ ladder in the \Omg\ phase with $\phi$=$1.829\pi$. The magnetization on each leg is almost identical.}
    \label{fig:omega_edge_modes}
\end{figure}
The most promising SPT candidate is the \Omg\ phase, where we have shown that $S^T_b$ is an active operator. It is then interesting to explicit
demonstrate the appearance of the edge states by applying a small field term of form $-h_b S^T_b $ to the ladder. However, since the states
are not eigenstates of $S^T_b$, this coupling is not simply a Zeeman term, although for small enough $h_b$ the change in energy should be linear in $h_b$. Hence, such a linear regime has to be located, and the field carefully applied within the linear regime. We select a field term
of $h_b=10^{-5}$, small enough that for $N$=200 the change in energy is significantly smaller than the gap in the system. Yet, this field is large enough that the very small finite splitting of the four states is irrelevant.
The resulting edge states are shown in Fig.~\ref{fig:omega_edge_modes} as obtained from finite DMRG calculations with OBC using cluster A
from Fig.~\ref{fig:ladder}. Here $n$ corresponds to the site index, with odd $n$ for the lower leg of the ladder and even $n$ for the upper.
Evidently, $\langle S^b_{n}\rangle$ is the same for both legs. Furthermore, the peak in $\langle S^b_{n}\rangle$ is not at sites 1, 2 but instead
occurs for sites $n$=5,6. The decay of the amplitude of  $\langle S^b_{n}\rangle$ is consistent with the previous estimate of the bulk correlation length of $\xi_\Omega\sim$30$a$, and we note that with the small field applied we find $\langle S^T_b\rangle\sim$3, slightly below the value of
3-4 estimated from the results in Fig.~\ref{fig:salpha} from much smaller systems with $h_b$=0.

\subsection{\Ups\ phase}
\begin{figure}
    \centering
    \includegraphics[scale=0.6]{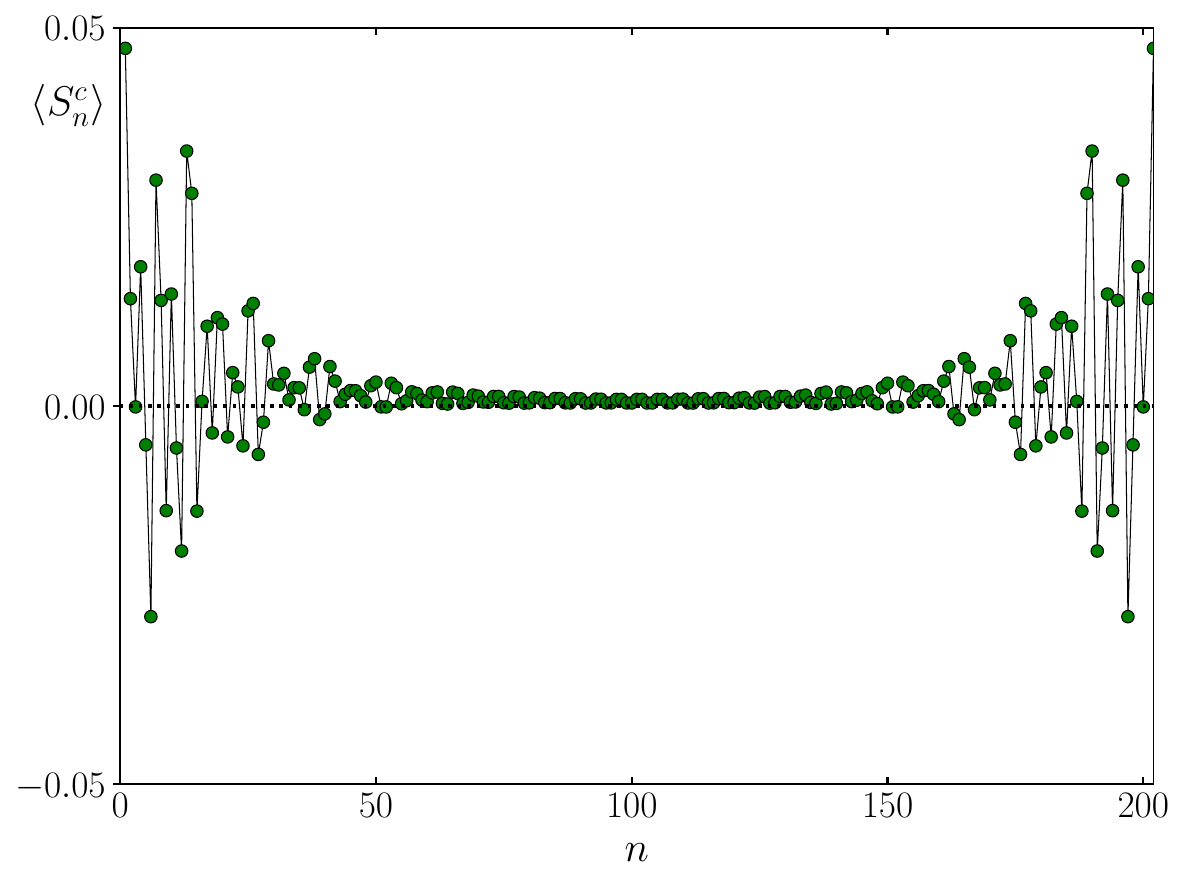}
    \caption{Magnetization $\langle S^c_{n}\rangle$ from finite DMRG calculations along the $c=[111]$ direction with a uniform field of 
    $|h_c|$=3$\times$10$^{-4}$ applied in the $[111]$ direction at every site.
    Results are shown for a $N=202$ ladder on cluster B with OBC, in the \Ups\ phase with $\phi$=$1.799\pi$. }
    \label{fig:upsilon_edge_modes}
\end{figure}
In a manner similar to the \Omg\ phase, we can study the edge-states appearing in the \Ups\ phase when the four degenerate ground-states are split
by a small field $h_c$, along the $c$ direction, $[111]$. We use a field strength of $|h_c|$=3$\times$10$^{-4}$ in the linear regime of the field.
The results are shown in Fig.~\ref{fig:upsilon_edge_modes} as obtained from finite DMRG calculations with OBC using cluster B from Fig.~\ref{fig:ladder}. Edge states at either end of the ladder are clearly visible. In contrast to the results for the \Omg\ phase, the two legs of the ladder do not show identical behavior. In fact, due to the shape of cluster B, $\langle S^c_{n}\rangle$ on the upper leg at the left half of the ladder is identical to the results on the {\it lower} leg on the right half of the ladder. As expected, the results in Fig.~\ref{fig:upsilon_edge_modes} show $S^T_c\sim 0.52$, consistent with our finding of $s_c\sim0.5$ in the \Ups\ phase.

\subsection{\dlt\ phase}
\begin{figure}
    \centering
  \includegraphics[scale=0.6]{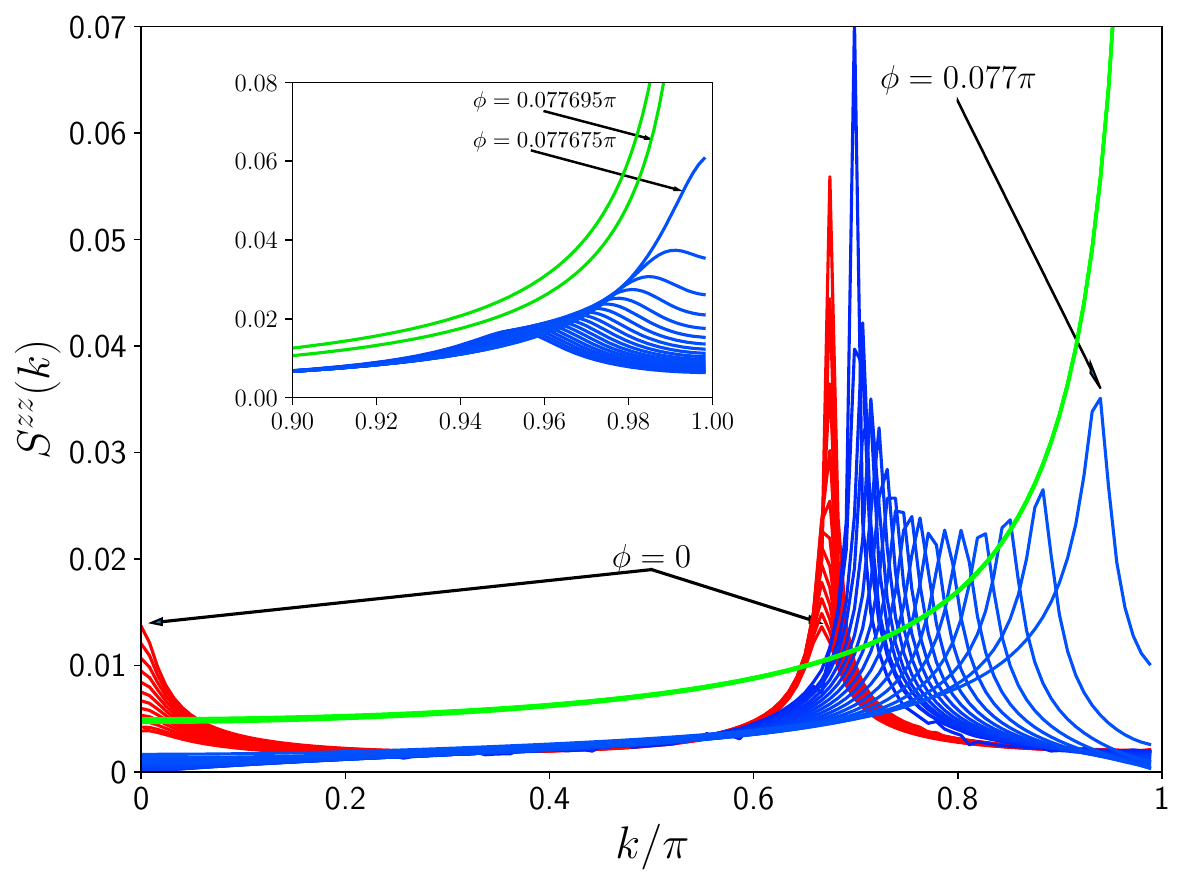}
  \caption{ Structure factor $S^{zz}(k)$ from the $\langle S^z_i S^z_{i+r} \rangle$ correlation functions of the AF-$\Gamma$, $\delta$ and AF phases obtained from iDMRG along the first leg of the ladder, with $r$ is measured along the leg. The red color indicates where $S^{zz}(k)$ is in the AF-$\Gamma$ phase, transitioning to the $\delta$ phase in blue, and ending in the AF phase in green. $S^{zz}(k)$ at $\phi=0$ is pointed out as having two peaks, one at $k=0$ and at $k\approx 2/3$. The last point in the $\delta$ phase in this sweep is also pointed out at $\phi=0.077\pi$ before the transition to the AF phase occurs. The inset is a higher precision calculation of $S^{zz}(k)$ near the $\delta$-AF transition. Starting at $\phi=0.077314\pi$ in the $\delta$ phase in blue, $\phi$ is increased to a maximum value of $\phi= 0.077716\pi$ in green in the AF phase. The last value in the $\delta$ phase occurs at $\phi = 0.077675\pi$ while the first value in the AF occurs at $\phi = 0.077695\pi$, both centered around $k = 1$.}\label{fig:fourier}
\end{figure}  
For the \dlt\ phase, we have not been able to obtain a clear picture of any eventual edge states. One reason for this is likely the very large correlation length, in excess of $\xi_\delta\sim$57$a$ throughout the phase. However, another important effect is the appearance of pronounced incommensurate
correlations, as we shall now discuss. The first thing we note is that from the results presented in Fig.~\ref{fig:phase_diagram} and Fig.~\ref{fig:xi}, it is clear that the correlation length diverges and the gap goes to zero at either end of the \dlt\ phase. The \dlt\ phase
is then a well-defined phase and not simply a part of either the \AG\ or \AF\ phases marked by the onset of incommensurate correlations. We can analyze the correlations by Fourier transforming the $\langle S^z_i S^z_{i+n} \rangle$ correlation functions along the first leg of the ladder. The resulting structure factors
$S^{zz}(k)$ are shown in Fig.~\ref{fig:fourier} for values of $\phi/ \pi$ starting in the \AG\ phase and ending in the \AF\ phase.
In the \AG\ phase at $\phi$=0 the correlations along a leg has a simple periodicity of 3, corresponding to a peak in the structure factor at $k$=2/3$\pi$. On the other hand, in the \AF\ phase, the peak $S^{zz}(k)$ must be at $k$=$\pi$. However, inside the \dlt\ phase, the peak in 
$S^{zz}(k)$ moves continuously from $k$=2/3$\pi$ to $k$=$\pi$. Close to the \dlt-\AF\ transition we show in the inset of Fig.~\ref{fig:fourier}
higher precision results for the behavior of $S^{zz}(k)$. To within our numerical precision, it appears that the peak value of $S^{zz}(k)$ does not
jump at the quantum critical points, but indeed moves continuously between the two limits.

\section{Conclusion}
Stranger things have happened, but the observation of eleven well-defined phases for the zero field phase diagram of the \JG-ladder is remarkable.
The proliferation of phases is due to the presence of the \G\ interaction term, which lowers the symmetry of the model, allowing for a finely tuned competition between the various phases. It would be of considerable interest to identify low-dimensional materials representative of this model. Presently,
we are not aware of any clear candidates. However,  the class of Kitaev materials is rapidly expanding, and it is thus plausible that materials with dominant
antiferromagnetic Heisenberg interactions and sub dominant \G-interactions can be found.

Among the eleven phases we have identified three new phases, the \Ups, \Omg\ and \dlt\ phases which do not show
signs of any ordinary long-range magnetic order and could potentially be SPT phases. We have also not found any
indication of valence bond ordering. However, we cannot rigorously rule out that the states can be reduced to trivial
product states on a large enough length scale. However, such a length scale would have to be sizable, and this scenario seems unlikely.
Among the three phases, the \Omg\ phase appears as the most likely SPT
phase and clear edge states are observed when a magnetic field is applied along the $[1\bar{1}0]$ direction. 
Similarly, for the \Ups\ phase, the application of a field along the $[111]$ direction induces clear edge states.

In future work, it would be fascinating to investigate the phase diagram of the \JG\ ladder in the presence of an applied field, which
we expect to show an abundance of new and intriguing phases.

\ack
We acknowledge the support of
the Natural Sciences and Engineering Research Council of Canada (NSERC) through Discovery
Grant No. RGPIN-2017-05759.
This research was enabled in part by support provided by SHARCNET (sharcnet.ca) and the Digital Research Alliance of Canada (alliancecan.ca).
Part of the numerical
calculations were performed using the ITensor library~\cite{itensor}.

\section*{References}
\bibliography{references}

\end{document}

%% file: tkz.tex
\usepackage{tikz}
\usetikzlibrary{arrows}
\usepackage{ifthen}
\usetikzlibrary{decorations.pathreplacing,decorations.markings}

\tikzset{
  on each segment/.style={
    decorate,
    decoration={
      show path construction,
      moveto code={},
      lineto code={
        \path [#1]
        (\tikzinputsegmentfirst) -- (\tikzinputsegmentlast);
      },
      curveto code={
        \path [#1] (\tikzinputsegmentfirst)
        .. controls
        (\tikzinputsegmentsupporta) and (\tikzinputsegmentsupportb)
        ..
        (\tikzinputsegmentlast);
      },
      closepath code={
        \path [#1]
        (\tikzinputsegmentfirst) -- (\tikzinputsegmentlast);
      },
    },
  },
  mid arrow/.style={postaction={decorate,decoration={
        markings,
        mark=at position .5 with {\arrow[#1]{stealth}}
      }}},
}

\def\circledarrow#1#2#3{ 
\draw[#1,->] (#2) +(80:#3) arc(80:-260:#3);
}
\def\rs{45}
\def\chrlAK#1{%
  \pgfmathparse{int(#1)}%
  \ifnum\pgfmathresult>0
  \foreach \x in {0,...,#1}
     \filldraw (1.3*\x,-.1) circle (0.1);
  \foreach \x in {0,...,#1}
     \filldraw (1.3*\x,1.1) circle (0.1);
  \draw[line width=0.2mm] (-0.2,1.1) -- (8.0,1.1);
  \draw[line width=0.2mm] (-0.2,-.1) -- (8.0,-.1);
   \pgfmathtruncatemacro{\y}{#1}
  \foreach \x in {0,...,\y}
        \draw[line width=0.2mm] (1.3*\x,1.1) -- (1.3*\x,-.1);
  \foreach \x in {0,...,5} {
     \ifthenelse{\x=0 \OR \x=2 \OR \x=4}{
     \path [draw=red,line width=\rs*0.0023882976987468mm,postaction={on each segment={mid arrow=red}}]
        (1.3*\x+0.1,0) -- (1.3*\x+1.1,0) -- (1.3*\x+0.1,1)-- (1.3*\x+0.1,0);
      \draw (1.3*\x+0.4,0.35) node {-};
      \path [draw=blue,line width=\rs*0.0023882976987468mm,postaction={on each segment={mid arrow=blue}}]
         (1.3*\x+0.2,1.0) -- (1.3*\x+1.2,1.00) -- (1.3*\x+1.2,0.00) -- (1.3*\x+0.2,1.0);
      \draw (1.3*\x+0.85,0.65) node {+};
    } {
     \path [draw=purple,line width=\rs*0.0058263516163867mm,postaction={on each segment={mid arrow=purple}}]
        (1.3*\x+0.1,0) -- (1.3*\x+1.1,0) -- (1.3*\x+0.1,1)-- (1.3*\x+0.1,0);
      \draw (1.3*\x+0.4,0.35) node {-};
      \path [draw=blue,line width=\rs*0.0058263516163867mm,postaction={on each segment={mid arrow=blue}}]
         (1.3*\x+0.2,1.0) -- (1.3*\x+1.2,1.00) -- (1.3*\x+1.2,0.00) -- (1.3*\x+0.2,1.0);
      \draw (1.3*\x+0.85,0.65) node {+};
      }
    }
    \fi
  }
  
\def\ra{15}
\def\chrlfmz#1{%
  \pgfmathparse{int(#1)}%
  \ifnum\pgfmathresult>0
  \foreach \x in {0,...,#1}
     \filldraw (1.3*\x,-.1) circle (0.1);
  \foreach \x in {0,...,#1}
     \filldraw (1.3*\x,1.1) circle (0.1);
  \draw[line width=0.2mm] (-0.2,1.1) -- (8.0,1.1);
  \draw[line width=0.2mm] (-0.2,-.1) -- (8.0,-.1);
   \pgfmathtruncatemacro{\y}{#1}
  \foreach \x in {0,...,\y}
        \draw[line width=0.2mm] (1.3*\x,1.1) -- (1.3*\x,-.1);
  \foreach \x in {0,...,5} {
     \ifthenelse{\x=0 \OR \x=2 \OR \x=4}{
     \path [draw=blue,line width=\ra*0.0013907549727983mm,postaction={on each segment={mid arrow=blue}}]
        (1.3*\x+0.1,0) -- (1.3*\x+0.1,1) -- (1.3*\x+1.1,0)-- (1.3*\x+0.1,0);
      \draw (1.3*\x+0.4,0.35) node {+};
      \path [draw=blue,line width=\ra*0.0013907549727983mm,postaction={on each segment={mid arrow=blue}}]
         (1.3*\x+0.2,1.0) -- (1.3*\x+1.2,1.00) -- (1.3*\x+1.2,0.00) -- (1.3*\x+0.2,1.0);
      \draw (1.3*\x+0.85,0.65) node {+};
    } {
     \path [draw=red,line width=\ra*0.0013907549727983mm,postaction={on each segment={mid arrow=red}}]
        (1.3*\x+0.1,0) -- (1.3*\x+1.1,0) -- (1.3*\x+0.1,1)-- (1.3*\x+0.1,0);
      \draw (1.3*\x+0.4,0.35) node {-};
      \path [draw=red,line width=\ra*0.0013907549727983mm,postaction={on each segment={mid arrow=red}}]
         (1.3*\x+0.2,1.0) -- (1.3*\x+1.2,0.00)-- (1.3*\x+1.2,1.00) -- (1.3*\x+0.2,1.0);
      \draw (1.3*\x+0.85,0.65) node {-};
      }
    }
    \fi
  }

\def\ra{15}
\def\chrlfmxy#1{%
  \pgfmathparse{int(#1)}%
  \ifnum\pgfmathresult>0
  \foreach \x in {0,...,#1}
     \filldraw (1.3*\x,-.1) circle (0.1);
  \foreach \x in {0,...,#1}
     \filldraw (1.3*\x,1.1) circle (0.1);
  \draw[line width=0.2mm] (-0.2,1.1) -- (8.0,1.1);
  \draw[line width=0.2mm] (-0.2,-.1) -- (8.0,-.1);
   \pgfmathtruncatemacro{\y}{#1}
  \foreach \x in {0,...,\y}
        \draw[line width=0.2mm] (1.3*\x,1.1) -- (1.3*\x,-.1);
  \foreach \x in {0,...,5} {
     \ifthenelse{\x=0 \OR \x=2 \OR \x=4}{
     \path [draw=blue,line width=\ra*0.0165864073812376mm,postaction={on each segment={mid arrow=blue}}]
        (1.3*\x+0.1,0) -- (1.3*\x+0.1,1) -- (1.3*\x+1.1,0)-- (1.3*\x+0.1,0);
      \draw (1.3*\x+0.4,0.35) node {+};
      \path [draw=blue,line width=\ra*0.016586407383458mm,postaction={on each segment={mid arrow=blue}}]
         (1.3*\x+0.2,1.0) -- (1.3*\x+1.2,1.00) -- (1.3*\x+1.2,0.00) -- (1.3*\x+0.2,1.0);
      \draw (1.3*\x+0.85,0.65) node {+};
    } {
     \path [draw=red,line width=\ra*0.016586407383458mm,postaction={on each segment={mid arrow=red}}]
        (1.3*\x+0.1,0) -- (1.3*\x+1.1,0) -- (1.3*\x+0.1,1)-- (1.3*\x+0.1,0);
      \draw (1.3*\x+0.4,0.35) node {-};
      \path [draw=red,line width=\ra*0.0165864073812376mm,postaction={on each segment={mid arrow=red}}]
         (1.3*\x+0.2,1.0) -- (1.3*\x+1.2,0.00)-- (1.3*\x+1.2,1.00) -- (1.3*\x+0.2,1.0);
      \draw (1.3*\x+0.85,0.65) node {-};
      }
    }
    \fi
  }

\def\ra{15}
\def\chrlalpha#1{%
  \pgfmathparse{int(#1)}%
  \ifnum\pgfmathresult>0
  \foreach \x in {0,...,#1}
     \filldraw (1.3*\x,-.1) circle (0.1);
  \foreach \x in {0,...,#1}
     \filldraw (1.3*\x,1.1) circle (0.1);
  \draw[line width=0.2mm] (-0.2,1.1) -- (8.0,1.1);
  \draw[line width=0.2mm] (-0.2,-.1) -- (8.0,-.1);
   \pgfmathtruncatemacro{\y}{#1}
  \foreach \x in {0,...,\y}
        \draw[line width=0.2mm] (1.3*\x,1.1) -- (1.3*\x,-.1);
  \foreach \x in {0,...,5} {
     \ifthenelse{\x=0 \OR \x=2 \OR \x=4}{
     \path [draw=blue,line width=\ra*0.0173986550528829mm,postaction={on each segment={mid arrow=blue}}]
        (1.3*\x+0.1,0) -- (1.3*\x+0.1,1) -- (1.3*\x+1.1,0)-- (1.3*\x+0.1,0);
      \draw (1.3*\x+0.4,0.35) node {+};
      \path [draw=red,line width=\ra*0.0173986550528829mm,postaction={on each segment={mid arrow=red}}]
         (1.3*\x+0.2,1.0) -- (1.3*\x+1.2,0.00)-- (1.3*\x+1.2,1.00) -- (1.3*\x+0.2,1.0);
      \draw (1.3*\x+0.85,0.65) node {-};
    } {
     \path [draw=red,line width=\ra*0.0173986550528829mm,postaction={on each segment={mid arrow=red}}]
        (1.3*\x+0.1,0) -- (1.3*\x+1.1,0) -- (1.3*\x+0.1,1)-- (1.3*\x+0.1,0);
      \draw (1.3*\x+0.4,0.35) node {-};
      \path [draw=blue,line width=\ra*0.0173986550528829mm,postaction={on each segment={mid arrow=blue}}]
         (1.3*\x+0.2,1.0) -- (1.3*\x+1.2,1.00) -- (1.3*\x+1.2,0.00) -- (1.3*\x+0.2,1.0);
      \draw (1.3*\x+0.85,0.65) node {+};
      }
    }
    \fi
  }

\def\chrldelta#1{%
  \pgfmathparse{int(#1)}%
  \ifnum\pgfmathresult>0
  \foreach \x in {0,...,#1}
     \filldraw (1.3*\x,-.1) circle (0.1);
  \foreach \x in {0,...,#1}
     \filldraw (1.3*\x,1.1) circle (0.1);
  \draw[line width=0.2mm] (-0.2,1.1) -- (8.0,1.1);
  \draw[line width=0.2mm] (-0.2,-.1) -- (8.0,-.1);
   \pgfmathtruncatemacro{\y}{#1}
  \foreach \x in {0,...,\y}
        \draw[line width=0.2mm] (1.3*\x,1.1) -- (1.3*\x,-.1);
  \foreach \x in {0,...,5} {
     \ifthenelse{\x=0 \OR \x=2 \OR \x=4}{
     \path [draw=red,line width=\ra*0.0254952100792158mm,postaction={on each segment={mid arrow=red}}]
        (1.3*\x+0.1,0) -- (1.3*\x+1.1,0) -- (1.3*\x+0.1,1)-- (1.3*\x+0.1,0);
      \draw (1.3*\x+0.4,0.35) node {-};
      \path [draw=purple,line width=\ra*0.0372949466940812mm,postaction={on each segment={mid arrow=purple}}]
         (1.3*\x+0.2,1.0) -- (1.3*\x+1.2,0.00) -- (1.3*\x+1.2,1.00) -- (1.3*\x+0.2,1.0);
      \draw (1.3*\x+0.85,0.65) node {-};
    } {
     \path [draw=purple,line width=\ra*0.0372949466940812mm,postaction={on each segment={mid arrow=purple}}]
        (1.3*\x+0.1,0) -- (1.3*\x+1.1,0) -- (1.3*\x+0.1,1)-- (1.3*\x+0.1,0);
      \draw (1.3*\x+0.4,0.35) node {-};
      \path [draw=red,line width=\ra*0.0254952100792158mm,postaction={on each segment={mid arrow=red}}]
         (1.3*\x+0.2,1.0) -- (1.3*\x+1.2,0.00) -- (1.3*\x+1.2,1.00) -- (1.3*\x+0.2,1.0);
      \draw (1.3*\x+0.85,0.65) node {-};
      }
    }
    \fi
  }

\def\kgusix#1{%
  \pgfmathparse{int(#1)}%
  \ifnum\pgfmathresult>0
  \foreach \x in {0,...,#1}
     \pgfmathtruncatemacro{\y}{2*\x+1}
     \draw (1.3*\x,-.1) node[circle,scale = 0.6, fill=black, black, draw, label=below:\y](b){};
  \foreach \x in {0,...,#1}
     \pgfmathtruncatemacro{\y}{2*\x+2}
     \draw (1.3*\x,1.1) node[circle,scale = 0.6, fill=black, black, draw, label=above:\y](b){};
  \draw[line width=0.2mm] (-0.2,1.1) -- (8.0,1.1);
  \draw[line width=0.2mm] (-0.2,-.1) -- (8.0,-.1);
   \pgfmathtruncatemacro{\y}{#1-1}
  \foreach \x in {0,...,\y} {
    \ifthenelse{\x=0 \OR \x=3 \OR \x=6}{
      \draw[line width=0.2mm] (1.3*\x,1.1) -- node[above] {$x'$} (1.3*\x+1.3,1.1);
      \draw[line width=0.2mm] (1.3*\x,-.1) -- node[below] {$x'$} (1.3*\x+1.3,-.1);
      } {
    \ifthenelse{\x=1 \OR \x=4 \OR \x=7}{
      \draw[line width=0.2mm] (1.3*\x,1.1) -- node[above] {$z'$} (1.3*\x+1.3,1.1);
      \draw[line width=0.2mm] (1.3*\x,-.1) -- node[below] {$z'$} (1.3*\x+1.3,-.1);
    }{
    \ifthenelse{\x=2 \OR \x=5 \OR \x=8}{
      \draw[line width=0.2mm] (1.3*\x,1.1) -- node[above] {$y'$} (1.3*\x+1.3,1.1);
      \draw[line width=0.2mm] (1.3*\x,-.1) -- node[below] {$y'$} (1.3*\x+1.3,-.1);
    }{}}
  }
  }
   \pgfmathtruncatemacro{\y}{#1}
  \foreach \x in {0,...,\y} {
    \ifthenelse{\x=0 \OR \x=3 \OR \x=6}{
      \draw[line width=0.2mm] (1.3*\x,1.1) -- node[left] {$z'$} (1.3*\x,-.1);
      } {
    \ifthenelse{\x=1 \OR \x=4 \OR \x=7}{
      \draw[line width=0.2mm] (1.3*\x,1.1) -- node[left] {$y'$} (1.3*\x,-.1);
    }{
    \ifthenelse{\x=2 \OR \x=5 \OR \x=8}{
      \draw[line width=0.2mm] (1.3*\x,1.1) -- node[left] {$x'$} (1.3*\x,-.1);
    }{}}
  }
  }
  \fi
  }

\def\wkgusixw#1{%
  \pgfmathparse{int(#1)}%
  \ifnum\pgfmathresult>0
  \foreach \x in {0,...,#1}
     \pgfmathtruncatemacro{\y}{2*\x+1}
     \draw (1.3*\x,-.1) node[circle,scale = 0.6, fill=black, black, draw, label=below:\y](b){};
  \foreach \x in {0,...,#1}
     \pgfmathtruncatemacro{\y}{2*\x+2}
     \draw (1.3*\x,1.1) node[circle,scale = 0.6, fill=black, black, draw, label=above:\y](b){};
  \draw[line width=0.2mm] (-0.2,1.1) -- (8.0,1.1);
  \draw[line width=0.2mm] (-0.2,-.1) -- (8.0,-.1);
   \pgfmathtruncatemacro{\y}{#1-1}
  \foreach \x in {0,...,\y} {
    \ifthenelse{\x=0 \OR \x=3 \OR \x=6}{
      \draw[line width=0.2mm] (1.3*\x,1.1) -- node[above] {$\Gamma_y\Gamma_z$} (1.3*\x+1.3,1.1);
      \draw[line width=0.2mm] (1.3*\x,-.1) -- node[below] {$\Gamma_y\Gamma_z$} (1.3*\x+1.3,-.1);
      \ifthenelse{\x=0}{
        \filldraw (1.3*\x+0.5,0.5) node (text) {$K^{k,l}_{i,j}$};
       \circledarrow{thick, black}{text}{0.37cm};
      }{
        \filldraw (1.3*\x+0.5,0.5) node (text) {$K^{k,l}_{i,j}$};
       \circledarrow{thick, black}{text}{0.37cm};
      }
      } {
    \ifthenelse{\x=1 \OR \x=4 \OR \x=7}{
      \draw[line width=0.2mm] (1.3*\x,1.1) -- node[above] {$K_z\Gamma_y$} (1.3*\x+1.3,1.1);
      \draw[line width=0.2mm] (1.3*\x,-.1) -- node[below] {$K_z\Gamma_y$} (1.3*\x+1.3,-.1);
      \filldraw (1.3*\x+0.5,0.5) node (text) {$\Gamma^{k,l}_{i,j}$};
      \circledarrow{thick, black}{text}{0.37cm};
    }{
    \ifthenelse{\x=2 \OR \x=5 \OR \x=8}{
      \draw[line width=0.2mm] (1.3*\x,1.1) -- node[above] {$K_y\Gamma_z$} (1.3*\x+1.3,1.1);
      \draw[line width=0.2mm] (1.3*\x,-.1) -- node[below] {$K_y\Gamma_z$} (1.3*\x+1.3,-.1);
      \filldraw (1.3*\x+0.5,0.5) node (text) {$\Gamma^{k,l}_{i,j}$};
      \circledarrow{thick, black}{text}{0.37cm};
    }{}}
  }
  }
   \pgfmathtruncatemacro{\y}{#1}
  \foreach \x in {0,...,\y} {
    \ifthenelse{\x=0 \OR \x=3 \OR \x=6}{
      \draw[line width=0.2mm] (1.3*\x,1.1) -- node[left] {$z'$} (1.3*\x,-.1);
      } {
    \ifthenelse{\x=1 \OR \x=4 \OR \x=7}{
      \draw[line width=0.2mm] (1.3*\x,1.1) -- node[left] {$y'$} (1.3*\x,-.1);
    }{
    \ifthenelse{\x=2 \OR \x=5 \OR \x=8}{
      \draw[line width=0.2mm] (1.3*\x,1.1) -- node[left] {$x'$} (1.3*\x,-.1);
    }{}}
  }
  }
  \fi
  }

\def\rungtripletA#1{%
  \pgfmathparse{int(#1)}%
  \ifnum\pgfmathresult>0
  \draw (-0.6,0.5) node (b){A};
  \foreach \x in {0,...,#1}
     \pgfmathtruncatemacro{\y}{2*\x+1}
     \draw (1.3*\x,-.1) node[circle,scale = 0.6, fill=gray, gray, draw](b){};
  \foreach \x in {0,...,#1}
     \pgfmathtruncatemacro{\y}{2*\x+2}
     \draw (1.3*\x,1.1) node[circle,scale = 0.6, fill=gray, gray, draw](b){};
  \draw[line width=0.2mm,gray] (-0.2,1.1) -- (8.0,1.1);
  \draw[line width=0.2mm,gray] (-0.2,-.1) -- (8.0,-.1);
   \pgfmathtruncatemacro{\y}{#1-1}
  \foreach \x in {0,...,\y} {
      \draw[line width=0.2mm,gray] (1.3*\x,1.1) -- (1.3*\x+1.3,1.1);
      \draw[line width=0.2mm,gray] (1.3*\x,-.1) -- (1.3*\x+1.3,-.1);
  }
  \pgfmathtruncatemacro{\y}{#1}
  \foreach \x in {0,...,\y} {
    \ifthenelse{\x=0 \OR \x=3 \OR \x=6}{
      \draw[line width=0.2mm,gray] (1.3*\x,1.1) --  (1.3*\x,-.1);
      } {
    \ifthenelse{\x=1 \OR \x=4 \OR \x=7}{
      \draw[line width=0.2mm,gray] (1.3*\x,1.1) -- (1.3*\x,-.1);
    }{
    \ifthenelse{\x=2 \OR \x=5 \OR \x=8}{
      \draw[line width=0.2mm,gray] (1.3*\x,1.1) --  (1.3*\x,-.1);
    }{}}
  }
  }
  \pgfmathtruncatemacro{\y}{#1}
  \foreach \x in {0,...,\y} {
    \ifthenelse{\x=0 \OR \x=1 \OR \x=6}{
        \filldraw (1.3*\x,0.5) node (text) {$t_x$};
        \filldraw[green, fill opacity=0.3] (1.3*\x,0.5) ellipse (0.24 and 0.9);
      }{}
    \ifthenelse{\x=2 \OR \x=3}{
        \filldraw (1.3*\x,0.5) node (text) {$t_y$};
        \filldraw[red, fill opacity=0.3] (1.3*\x,0.5) ellipse (0.24 and 0.9);
      }{}
    \ifthenelse{\x=4 \OR \x=5}{
        \filldraw (1.3*\x,0.5) node (text) {$t_z$};
        \filldraw[blue, fill opacity=0.3] (1.3*\x,0.5) ellipse (0.24 and 0.9);
      }{}
  }
  \fi
  }

\def\rungtripletB#1{%
  \pgfmathparse{int(#1)}%
  \ifnum\pgfmathresult>0
  \draw (-0.6,0.5) node (b){B};
  \foreach \x in {0,...,#1}
     \pgfmathtruncatemacro{\y}{2*\x+1}
     \draw (1.3*\x,-.1) node[circle,scale = 0.6, fill=gray, gray, draw](b){};
  \foreach \x in {0,...,#1}
     \pgfmathtruncatemacro{\y}{2*\x+2}
     \draw (1.3*\x,1.1) node[circle,scale = 0.6, fill=gray, gray, draw](b){};
  \draw[line width=0.2mm,gray] (-0.2,1.1) -- (8.0,1.1);
  \draw[line width=0.2mm,gray] (-0.2,-.1) -- (8.0,-.1);
   \pgfmathtruncatemacro{\y}{#1-1}
  \foreach \x in {0,...,\y} {
      \draw[line width=0.2mm,gray] (1.3*\x,1.1) -- (1.3*\x+1.3,1.1);
      \draw[line width=0.2mm,gray] (1.3*\x,-.1) -- (1.3*\x+1.3,-.1);
  }
  \pgfmathtruncatemacro{\y}{#1}
  \foreach \x in {0,...,\y} {
    \ifthenelse{\x=0 \OR \x=3 \OR \x=6}{
      \draw[line width=0.2mm,gray] (1.3*\x,1.1) --  (1.3*\x,-.1);
      } {
    \ifthenelse{\x=1 \OR \x=4 \OR \x=7}{
      \draw[line width=0.2mm,gray] (1.3*\x,1.1) -- (1.3*\x,-.1);
    }{
    \ifthenelse{\x=2 \OR \x=5 \OR \x=8}{
      \draw[line width=0.2mm,gray] (1.3*\x,1.1) --  (1.3*\x,-.1);
    }{}}
  }
  }
  \pgfmathtruncatemacro{\y}{#1}
  \foreach \x in {0,...,\y} {
    \ifthenelse{\x=0 \OR \x=5 \OR \x=6}{
        \filldraw (1.3*\x,0.5) node (text) {$t_y$};
        \filldraw[red, fill opacity=0.3] (1.3*\x,0.5) ellipse (0.24 and 0.9);
      }{}
    \ifthenelse{\x=3 \OR \x=4}{
        \filldraw (1.3*\x,0.5) node (text) {$t_x$};
        \filldraw[green, fill opacity=0.3] (1.3*\x,0.5) ellipse (0.24 and 0.9);
      }{}
    \ifthenelse{\x=1 \OR \x=2}{
        \filldraw (1.3*\x,0.5) node (text) {$t_z$};
        \filldraw[blue, fill opacity=0.3] (1.3*\x,0.5) ellipse (0.24 and 0.9);
      }{}
  }
  \fi
  }
